\documentclass{aa}  

\usepackage{graphicx}
\usepackage{soul,censor}
\newlength\nextcharwidth
\makeatletter
\renewcommand\@cenword[1]{%
  \setlength{\nextcharwidth}{\widthof{#1}}%
  \censorrule{\nextcharwidth}%
  \kern -\nextcharwidth%
  #1}
\makeatother

\usepackage{txfonts}
\usepackage[colorlinks = true,
linkcolor= blue,
citecolor = blue,
filecolor = blue,
urlcolor = blue,
]{hyperref}

\defcitealias{rebassa-mansergas2021}{RM+2021}
\usepackage{booktabs,dcolumn}
\newcolumntype{P}[1]{>{\centering\arraybackslash}p{#1}}

\begin{document}

   \title{Kinematic properties of white dwarfs}

   \subtitle{Galactic orbital parameters and age-velocity dispersion relation\thanks{Full tables\,2, 4--7 are available in electronic form
at the CDS via anonymous ftp to \href{http://cdsarc.u-strasbg.fr}{cdsarc.u-strasbg.fr} (\href{ftp://130.79.128.5}{\texttt{130.79.128.5}})
or via \url{http://cdsarc.u-strasbg.fr/viz-bin/cat/J/A+A/658/A22}}}

   \author{Roberto Raddi
          \inst{1}
          \and Santiago Torres \inst{1,2}          
          \and Alberto Rebassa-Mansergas \inst{1,2}
          \and Jes\'us Maldonado \inst{3}
          \and Mar\'ia E. Camisassa \inst{4}
          \and \\
          Detlev Koester \inst{5}
          \and Nicola Pietro Gentile Fusillo
          \inst{6}
          \and
          Pier-Emmanuel Tremblay
          \inst{7}
          \and
          Markus Dimpel
          \inst{8}
          \and
          Ulrich Heber
          \inst{8}
          \and\\
          Tim Cunningham
          \inst{7}   
          \and
          Juan-Juan Ren
          \inst{9}
          }

   \institute{Universitat Polit\`ecnica de Catalunya, Departament de F\'isica, c/ Esteve Terrades 5, 08860 Castelldefels, Spain\\
   \email{roberto.raddi@upc.edu}
   \and
   Institut d'Estudis Espacials de Catalunya, Ed. Nexus-201, c/Gran Capit\'a 2-4, 08034 Barcelona, Spain
   \and
   INAF - Osservatorio Astronomico di Palermo, Piazza del Parlamento 1, 90134 Palermo, Italy
   \and
   Department of Applied Mathematics, University of Colorado, Boulder, CO, 80309-0526, USA
   \and 
   Institut f\"ur Theoretische Physik und Astrophysik, Christian-Albrechts-Universit\"at, Kiel 24118, Germany
   \and
   European
 Southern Observatory, Karl Schwarzschild Stra{\ss}e 2, Garching, 85748, Germany
   \and
   Department of Physics, University of Warwick, Coventry CV4 7AL, UK
   \and
   Dr. Karl Remeis-Observatory \& ECAP, Astronomical Institute, Friedrich-Alexander University Erlangen-Nuremberg (FAU), Sternwartstr. 7, 96049, Bamberg, Germany
   \and
   CAS Key Laboratory of Space Astronomy and Technology, National Astronomical Observatories, Chinese Academy of Sciences, Beijing 100101, People's Republic of China}

   \date{Received 21 July 2021 / Accepted 29 October 2021}

% \abstract{}{}{}{}{} 
% 5 {} token are mandatory
 
  \abstract
  % context heading (optional)
   {Both kinematic and chemical tagging of stellar populations have revealed a great deal of information on the past and recent history of the Milky Way, including its formation history, merger events,  and mixing of populations across the Galactic disk and halo.} % leave it empty if necessary  
  % aims heading (mandatory)
   {We present the first detailed 3D kinematic analysis of a sample of 3133 white dwarfs that use {\em Gaia} astrometry plus radial velocities, which were measured either by {\em Gaia} or by ground based spectroscopic observations. The sample includes either isolated white dwarfs that have direct radial velocity measurements, or white dwarfs that belong to common proper motion pairs that contain nondegenerate companions with  available radial velocities. A subset of common proper motion pairs also have metal abundances that have been measured by large scale spectroscopic surveys or by our own follow-up observations.}
  % methods heading (mandatory)
   {We used the white dwarfs as astrophysical clocks, by determining their masses and total ages via interpolation with dedicated evolutionary models. We also used the nondegenerate companions in common proper motions to chemically tag of the population. Combining  accurate radial velocities with {\em Gaia} astrometry and proper motions, we derived the velocity components of our sample in the Galactic rest frame and their Galactic orbital parameters.}
  % results heading (mandatory)
   {The sample is mostly located within $\sim$300\,pc from the Sun. It contains predominantly (90--95\,\%) thin-disk stars with close-to-circular Galactic orbits, while the remaining 5--10\,\% of stars have more eccentric trajectories and belong to the thick disk. We identified seven isolated white dwarfs and two common proper motion pairs as halo members. We determined the age -- velocity dispersion relation for the thin-disk members, which agrees with previous results that were achieved from different white dwarf samples without published radial velocities. The age -- velocity dispersion relation shows signatures of dynamical heating and saturation after 4--6\,Gyr. We observed a mild anti-correlation between [Fe/H] and the radial component of the average -- velocity dispersion, showing that dynamical mixing of populations takes place in the Galactic disk, as was detected via the analysis of other samples of FGK stars.}
  % conclusions heading (optional), leave it empty if necessary 
  {We have shown that a white dwarf sample with accurate 3D kinematics and well-measured chemical compositions enables a wider understanding of their population in the solar neighborhood and its connection with the Galactic chemodynamics. The legacy of existing spectroscopic surveys will be boosted by the availability of upcoming larger samples of white dwarfs and common proper motion pairs with more uniform high-quality data. }

   \titlerunning{Kinematic properties of white dwarfs}
\authorrunning{Roberto Raddi et al.}
   \maketitle

%
%-------------------------------------------------------------------

\section{Introduction}
Our Galaxy, the Milky Way, is a dynamic ensemble of stars, gas, dust, and dark matter. Its complexity is enhanced by the mixing  of its components through the Galactic potential \citep[][]{selwood2002} or by recent and past interactions with satellite galaxies \citep[][]{antoja2018, helmi2018, belokurov2018}. In the current era of large spectroscopic surveys, such as the Radial Velocity Experiment \citep[RAVE;][]{rave}, the Large Sky Area Multi-Object Fiber Spectroscopic Telescope \citep[LAMOST;][]{lamost}, the Galactic Archaeology with HERMES \citep[GALAH;][]{galah}, the {\em Gaia}-ESO survey \citep{gaia-eso}, and the Apache Point Observatory Galactic Evolution Experiment \citep[APOGEE;][]{apogee}, our advantage point from the inside of the Milky Way  favors a direct analysis of stellar populations. This favorable perspective has been enabling key understanding of the correlations among location, kinematics, metallicity, and stellar ages, via the analysis of hundreds to several thousands spectra \citep[][and references therein]{bland-hawthorn2016}. These large spectroscopic databases have contributed to identifying metallicity gradients along the galactocentric radius and vertical height above the plane, or detailed information on stellar orbits in the Galactic  disk and halo \citep[][]{boeche13,bergmann2014,hayden2020}. With increasing distance from the plane, increasing numbers of metal-poor stars are found that also posses enhanced abundances of $\alpha$-elements (Mg, Si, Ca, Ti) with respect to Fe. This leads to the commonly adopted distinction in thin and thick components of the Galactic disk \citep{hayden2015,bland-hawthorn2019}. The observed differences in abundance patterns of Galactic populations are interpreted as the signature of  sequential enrichment of the interstellar medium with the ejecta of core-collapse and thermonuclear supernovae, which occur with different timescales  and leave their traces in the atmospheres of present-day stars \citep{matteucci1986}. Even the solar neighborhood contains mix of stars that have likely formed in situ or migrated from the inner Galaxy \citep[][]{navarro2011, adibekyan2011}. The kinematic analysis of metal-rich or metal-poor and $\alpha$-enhanced or -depleted populations indicates that different mechanisms take place in the Galactic disk, such as secular heating   \citep[i.e., an increased velocity dispersion in the radial or vertical directions;][]{dehnen1998,aumer2009, aumer2016} and radial mixing \citep[migration and blurring of Galactic orbits for stars that were born at different galactocentric radii;][]{schonrich2009, minchev2018, hayden2018}. 
\begin{table*}
    \centering
        \caption{Selection criteria adopted for the identification of white dwarf candidates and common proper motion companions. }
    \begin{tabular}{@{}c@{}}
\toprule
\toprule
Quality cuts applied to white dwarfs and common proper motion companions\\
\midrule
$\varpi > 0$ AND $\sigma_\varpi/\varpi \leq 0.2$ AND \verb|visibility_periods_used|~$> 10$\\
\verb|phot_bp_mean_mag|~$< 20.5$ AND \verb|phot_bp_mean_flux_over_error|~$>10$ \\
\verb|phot_rp_mean_mag|~$< 20$ AND \verb|phot_rp_mean_flux_over_error|~$>10$ \\
$\beta < 0.1$ AND $|C*| \leq 5\, \sigma_{C^{*}}(G) $ AND fidelity~$>0.5$ AND \verb|ruwe|\,$<$\,max($\overline{\texttt{ruwe}}$\,+\,$\sigma_{\texttt{ruwe}}$, 1.4)\\
    \bottomrule
White dwarf color-magnitude cuts \\
    \midrule
$G+5-5\times\log{(1000/\varpi)}>9\times(G-G_{\rm RP})-3\times (G-G_{\rm RP})^{2}-0.9\times(G-G_{\rm RP})^{3}+(G-G_{\rm RP})^{4}+8.5$\\
$G+5-5\times\log{(1000/\varpi)}>6\times(G_{\rm BP}-G_{\rm RP})-0.9\times(G_{\rm BP}-G_{\rm RP})^{2}-0.9\times(G_{\rm BP}-G_{\rm RP})^{3}+0.35\times(G_{\rm BP}-G_{\rm RP})^{4}+8.5$\\
$G-G_{\rm RP}<0.9$ AND $G_{\rm BP}-G_{\rm RP}<1.9$\\
\bottomrule
\multicolumn{1}{l}{{\bf Notes:} $\beta$ and $C^*$ are defined by \citet{riello2021}.}
    \end{tabular}
    \label{t:gaia}
\end{table*}

In this dynamic context, white dwarfs can play a prominent role because they are the end products of low- to intermediate-mass stars \citep[$<10$\,M$_{\odot}$;][]{doherty2017,cummings2018} and are ubiquitous in the  stellar populations of all ages in the Galactic disk and halo, even 100\,pc away from the Sun \citep{torres2021}. Although white dwarfs can potentially trace the kinematics of both the youngest and oldest stellar population because of their billion year long cooling times and their reliability as astrophysical clocks \citep{althaus2010},  their use  for such studies is typically hindered by some of their physical properties. First, white dwarfs only ``superficially'' lose the memory of the chemical make up of their progenitors because all the elements that are heavier than hydrogen or helium sink in their interiors as a result of their strong gravitational field \citep[][]{schatzman1945}. Second, measuring precise and accurate radial velocities of white dwarfs from their pressure-broadened lines is a difficult endeavor that is complicated by the intrinsic redshift of spectral lines due to the gravitational field of white dwarfs \citep[e.g.][]{greenstein1977,shipman1977}. Various attempts have been made to overcome the problem of the missing radial velocity in various ways  \citep[][]{oppenheimer2001,reid2001,sion2014,rowell2019,torres2021}. Some of the earliest works that included radial velocities of hydrogen-atmosphere (DA spectral type) white dwarfs contained small samples of stars that were used to determine the average gravitational redshift and the dispersion of the Galactic velocity components \citep{trimble1972,wegner1974}. Another way to study the kinematics of white dwarfs relies on studying those that belong to common proper motion pairs because the systemic radial velocity can be more easily estimated from the nondegenerate companions \citep{wegner1981,silvestri2001}. However, as a result of the lack of accurate radial velocities, one of the largest sample of white dwarfs that satisfies these requirements came from the ESO supernova type Ia progenitor survey \citep[SPY;][]{napiwotzki2020}. Out of this sample, 634 DA white dwarfs were categorized in thin and thick disk or halo members in an attempt to estimate the white dwarf contribution to baryonic dark matter \citep[][]{pauli2006, richter2007,dimpel2018}. More recently, a larger sample of 20\,247 DA white dwarfs from the Sloan Digital Sky Survey (SDSS) Data Release 12 was thoroughly investigated by \citet{anguiano2017}, who measured their radial velocities. These authors detected the existence of a mean Galactic radial velocity gradient and an additional source of dynamical heating  from the analysis of the age -- velocity dispersion relation, with respect to the prediction for the secular evolution of the Galactic disk. Later attempts at finding correlations among white dwarf masses, kinematics, and ages found that the most massive white dwarfs have the smallest velocity dispersion, thus appearing to have experienced less dynamical interactions because their lifetimes are shorter \citep{wegg2012}. More recent work by \citet{cheng2020} has suggested that the kinematics of massive white dwarfs may carry the imprint of past binary mergers. A significant fraction of massive white dwarfs have a larger velocity dispersion according to this and are therefore older than they appear.

In this paper, we expand the previous kinematic analysis of isolated white dwarfs that we complement with common proper motion pairs that contain white dwarfs and nondegenerate companions that have radial velocity and metallicity measurements. Combining radial velocities with the astrometric data of the European Space Agency {\em Gaia} mission \citep[][]{gaia2016}  for the first time, we analyze the full 3D kinematics and derive the Galactic orbital parameters of the sample of isolated white dwarfs and common proper motion pairs. Hence, we analyze their correlations with the white dwarf ages and the chemical composition of their nondegenerate companions. We also discuss a few peculiar systems. Finally, we determine the age -- velocity dispersion for these white dwarfs by comparing our results to previous measurements.

%--------------------------------------------------------------------

\section{Sample selection}
\label{sec:sample}
\subsection{Isolated white dwarfs with radial velocity measurements}
\label{sec:sample_wd}
\begin{figure}
    \centering
    \includegraphics{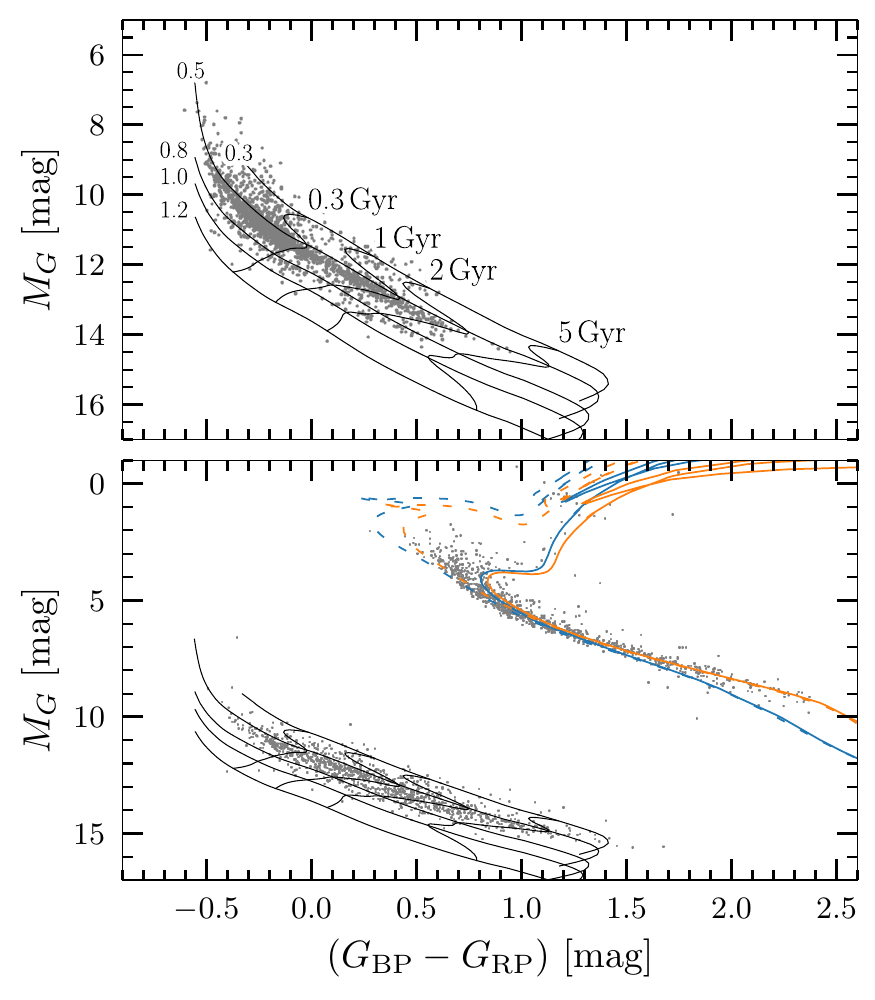}
    \caption{Hertzsprung-Russel diagrams of our selected white dwarfs with radial velocity measurements (top panel) and the {\em Gaia}-selected common proper motion pairs (bottom panel). The cooling tracks for \mbox{He-,} CO-, and ONe-core white dwarfs  \citep{althaus2013,camisassa2016,camisassa2019} and the isochrones are shown. The the corresponding masses and ages are labeled in the top panel, for clarity. The solar metallicity BaSTI isochrones with [Fe/H]~$=0.06$ \citep[blue curves;][]{hidalgo2018} and $\alpha$-enhancement \citep[orange curves;][]{pietrinferni2021} are plotted at 1 and 10~Gyr (dashed and solid curves, respectively).}
    \label{fig:gaia}
\end{figure}
Our first sample consisted of single white dwarfs that have radial velocity measurements. We cross-matched the latest release of SPY \citep[][]{napiwotzki2020}, which contains 643 DA white dwarfs, and the sample of 20\,247 DA white dwarfs from SDSS \citep[][]{anguiano2017} with the Early Data Release 3 of {\em Gaia} \citep[EDR3;][]{gaia2021}. We performed our cross-match via the TAP\footnote{Accessing the relevant online services, which are listed here:\\ \url{https://gea.esac.esa.int/tap-server/tap}\\
\url{https://gaia.ari.uni-heidelberg.de/tap}\\ \url{http://dc.zah.uni-heidelberg.de/tap}} service implemented by the command-line tool, \textsc{stilts} \citep{taylor2006}, requesting that parallaxes are strictly positive and are measured with a precision of 20\,\% ($\varpi >0$ and $\sigma_\varpi/\varpi<0.2$), and imposing additional quality cuts that are listed in Table\,\ref{t:gaia}.  Moreover, we cleaned our selection by requesting that our sources are minimally affected by blending, that is, they have the parameter $\beta < 0.1$ \citep[as defined by][]{riello2021}. We removed the most likely spurious astrometric solutions \citep[fidelity~$> 0.5$;][]{rybizki2021}, and imposed a 5$\sigma$ cut on the corrected BP and RP flux excess \citep[$C^{*}$;][]{riello2021}. As an additional safety measure, we also applied a cut on the renormalized unit weight error \citep[{\texttt{ruwe};}][]{lindegren2018-b} as defined in Table\,\ref{t:gaia}. Finally, we collected  the  geometric  distance  estimates  \citep[][]{bailer-jones2021} for all identified objects. In addition, we removed from the cross-match all the SPY white dwarfs  that either are identified as spectroscopic binaries, have other peculiarities, or have problematic data; we also excluded the SDSS white dwarfs with signal-to-noise ratio (S/N) lower than 20 and radial velocity errors larger than 10\,km\,s$^{-1}$. The result of this cross-match produced 484 white dwarfs with SPY spectra and 1641 white dwarfs with SDSS DR12 spectra; 48 stars are in common in the two surveys. As shown in the {\em Gaia} Hertzsprung-Russell (HR) diagram (top panel of Fig\,\ref{fig:gaia}), the selected objects encompass the full range of white dwarf masses and a wide spread of cooling ages up to 5 Gyr, although the majority of them are intrinsically bright white dwarfs younger than 0.3 Gyr. The {\em Gaia} source identifiers, photometry, and geometric distances of the selected white dwarfs are listed in Table\,\ref{tab:gaia}. The full table is available at the CDS.

\begin{figure}
    \centering
    \includegraphics{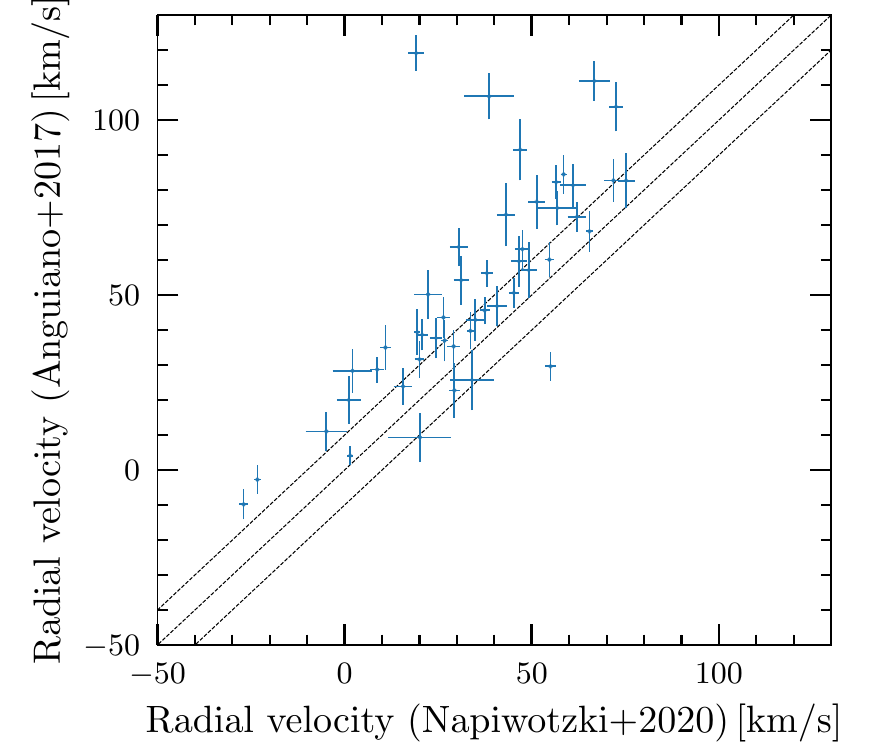}
    \caption{Radial velocity comparison for 48 isolated white dwarfs that have been observed by \citet{anguiano2017} and \citet{napiwotzki2020}. The dashed lines correspond to the equality and a $\pm 10$\,km\,s$^{-1}$ radial velocity offset.}
    \label{fig:rv_comparison_wd}
\end{figure}

Because it is important for our kinematic study, we note that radial velocities of the SPY and SDSS white dwarfs were measured with two different techniques. On the one hand, \citet{napiwotzki2020} performed Gaussian fits to the narrow  H$\alpha$ and H$\beta$ line cores of SPY white dwarfs \citep[a description of the procedure is detailed in Sect. 3 and Appendix A of][]{napiwotzki2020}, with the goal of avoiding the pressure shifts and asymmetries due to nonlinear effects in the wings of the strongly Stark-broadened Balmer lines  \citep[][and references therein]{halenka2015}. On the other hand, \citet{anguiano2017} used the cross-correlation method employing the entire spectra of SDSS white dwarfs. The two samples contain 48 white dwarfs in common. We compare the radial velocities in Fig\,\ref{fig:rv_comparison_wd}. The SDSS measurements show an average difference of $+15\pm13$\,km\,s$^{-1}$ with respect to the SPY results. Because the high-resolution SPY sample is better suited for accurate and precise radial velocity measurements, we corrected the SDSS sample for the average offset, and we used the SPY measurements when a star was observed by both surveys. 
\begin{table*}
    \centering
        \caption{System identifiers in this paper, {\em Gaia} identifiers, photometry, and distances of isolated white dwarfs and common proper motion pairs. }
 \scriptsize
    \begin{tabular}{@{}l
    l 
    c
    c 
    c 
    c
    l 
    c
    c 
    c 
    c@{}} 
    \toprule
    \toprule
  System &  
  {\em Gaia} EDR3  
  & $G$\,[mag]
  & $G_{\rm BP}$\,[mag] 
  & $G_{\rm RP}$\,[mag] 
  & $d$\,[pc] 
  & {\em Gaia} EDR3  
  & $G$\,[mag] 
  & $G_{\rm BP}$\,[mag]
  & $G_{\rm RP}$\,[mag]
  & $d$\,[pc]    \\
    \midrule   
\multicolumn{6}{c}{White dwarfs} & \multicolumn{5}{c}{nondegenerate companions} \\
         \midrule
0001 & 152935195517952 & $17.98$  & $18.00$ & $18.00$ & 225 &    &    &   &   &   \\
0002 & 288175125714560 & $17.47$  & $17.49$ & $17.52$ & 153 &    &    &   &   &   \\
.    & & & & & & & & & & \\
3148 & 6907694553263596672 & $19.54$  & $19.68$ & $19.39$ & 206 &  6907694690702552576 & $13.52$  & $14.11$ & $12.79$ & 223 \\
3149 & 6910805342238827648 & $16.37$  & $16.28$ & $16.59$ & 158 &  6910806102448648576 & $14.85$  & $15.88$ & $13.84$ & 156   \\
3150 & 6910816513450124288 & $17.40$  & $17.30$ & $17.60$ & 357 &    &    &   &   &   \\
3151 & 6917473674103954560 & $17.53$  & $17.69$ & $17.24$ & 75 &    &    &   &   &   \\
    \bottomrule
\multicolumn{11}{l}{{\bf Notes} The full table is available at the CDS. The system number is used as short identifier throughout the paper.}\\
    \end{tabular}
    \label{tab:gaia}
\end{table*}

\subsection{Common proper motion pairs}
\label{sec:sample_cpmp}

The second sample consisted of common proper motion pairs containing white dwarfs and nondegenerate companions that have radial velocity measurements. In order to identify the common proper motion pairs, we began by selecting white dwarf candidates following the methods of \citet{gentilefusillo2021}, and applying the color- and magnitude-cuts in proximity of the white dwarf cooling sequence in the {\em Gaia} EDR3 HR diagram as well as the quality cuts given in Table\,\ref{t:gaia}. The resulting query consisted of 168\,247 white dwarf candidates.

The identification of common proper motion pairs drew from the recent work of \citet{elbadry2021}, but we used our initial selection of white dwarfs for identifying mutual pairs and nondegenerate companions in the full \emph{Gaia} EDR3. In contrast with \citet{elbadry2021}, we searched for common proper motion pairs with maximum projected separations of $s \leq0.5 \times 10^{6}$\,au and a projected orbital-velocity difference of $\Delta\varv_{\rm orbit} \leq 2.7\times (s\,[{\rm au}]/1000)^{-1/2}$\,km\,s$^{-1}$, as it is computed for a total binary mass of $\sim$8.5\,M$_{\odot}$. We applied the quality cuts of Table\,\ref{t:gaia} to the selection of common proper motion companions, and we cleaned the resulting sample from possible members of open clusters by removing the stars that match the astrometric parameters of those characterized by \citet{cantat-gaudin2020}. The resulting sample contains 7256 systems, 40 of which are resolved triple systems containing two nondegenerate stars and six are triplets containing two white dwarfs. These systems are excluded from further analysis because they may have had higher chances of past interactions that affected their evolution \citep{toonen2020}. As for the single white dwarf sample, we collected the geometric distance estimates for all the identified pairs.

%--------------------------------------------------------------------
\subsection{Companion stars with radial velocity measurements}
Among the common proper motion pairs, we identified 976 systems for which the nondegenerate stars have reliable velocity measurements in the Second Data Release of {\em Gaia} \citep[DR2;][]{gaia2018}, based on their number of measurements  (\verb|dr2_rv_nb_transits > 5|) and the absence of brighter neighbors within 10\,arcsec \citep[reflecting stricter quality criteria than advised by][]{boubert2019}. This sample includes two double-degenerate binaries with a nondegenerate companion each, and one system containing one white dwarf and two nondegenerate companions that have very similar radial velocities. 

Subsequently, we searched for nondegenerate companion stars that may also have been observed by other spectroscopic surveys by cross-matching our sample with the RAVE DR5 \citep[][]{kunder2017}, LAMOST DR5 \citep[][]{luo2019,xiang2019}, APOGEE DR16 \citep[][]{jonsson2020}, and GALAH+ DR3 \citep[][]{buder2021}. Hence, we selected only those stars having iron abundances, [Fe/H], which is a proxy for stellar metallicity, and $\alpha$-element abundances, [$\alpha$/Fe], which is a measure of the chemical nucleosynthetic evolution. The selected data satisfy the quality criteria defined in the relevant references; when more than one spectrum was available for a star in given survey, we chose the best quality observation. The correlation between [Fe/H] and [$\alpha$/Fe] of this sample will be discussed in Sect.\,\ref{sec:discussion} because it is relevant for the interpretation of the results. In Table\,\ref{tab:spec-surveys} we list the numbers of spectra identified in each survey and the corresponding selection criteria. Finally, we added 61 common proper motion pairs to this sample that have spectra for their nondegenerate companions that were analyzed by \citet{rebassa-mansergas2021}, hereafter shortened to RM+2021 in figures and tables. The selected stars have spectra with a S/N~$>20$ and [Fe/H] measurements. For 48 of these stars we measured radial velocities, and the remaining 13 have reliable {\em Gaia} DR2 measurements. 
\begin{table}
\small
    \centering
    \caption{Results of our cross-match with spectroscopic surveys for the nondegenerate companions.}    
    \begin{tabular}{@{}lcc@{}}
     \toprule
     \toprule
      Catalogu & Quality criteria & Selected  \\
     \midrule  
     {\em Gaia} & \verb|dr2_rv_nb_transits| > 5, no bright neighbor  & 976  \\
     RAVE     & \verb|QK| = 0, \verb|c1 = c2 = c3 = n| & 68   \\
    LAMOST     & S/N > 20 & 113 \\    
    APOGEE     & \verb|Eflag| = 0, \verb|Aflag| = 0 & 37 \\   
    GALAH+     & \verb|flag_sp| = 0, \verb|flag_fe_h| = 0 & 51 \\
    \bottomrule
    \end{tabular}
    \label{tab:spec-surveys}
\end{table}
\begin{table*}
    \centering
        \caption{Element abundances and radial velocities of the {\em Gaia}-selected common proper motion pairs.}

    \begin{tabular}{@{}@{}P{1.0cm}@{}
    c
    c
    c
    D{,}{\,\pm\,}{9}
    D{,}{\,\pm\,}{5}} 
    \toprule
  System 
  & Survey  
  & \multicolumn{1}{c}{[Fe/H]} 
  & \multicolumn{1}{c}{[$\alpha$/Fe]} 
  & \multicolumn{1}{c}{$\varv_{\rm rad}$\,[km\,s$^{-1}$]$^{(1)}$} 
  & \multicolumn{1}{c}{$\varv_{\rm rad}$\,[km\,s$^{-1}$]$^{(2)}$} \\
    \midrule 
0011  & CPMP&  	      & 	       &		 & +32.3,1.7 \\
0020  & GALAH+& $+0.03\pm0.10$ & $+0.06\pm0.07$ &  -11.3,0.1 &		  \\
. & & & & \\
3146  & RAVE& $-0.22\pm0.20$ & $+0.14\pm0.20$ &  -23.9,0.9 & -24.2,0.3 \\
3148  & CPMP&  	      & 	       &		 & +32.3,4.9 \\
         \bottomrule
\multicolumn{5}{l}{{\bf Notes.} The full table is available at the CDS.}\\
\multicolumn{5}{l}{CPMP indicates the nondegenerate stars that do not have available spectra.}\\
\multicolumn{6}{l}{The radial velocities from (1) the listed surveys or (2) {\em Gaia} DR2.}\\
    \end{tabular}
    \label{tab:cpmp}
\end{table*}

All in all, our sample of common proper motion pairs with radial velocity measurements includes 1092 systems, which are displayed in the \emph{Gaia} HR diagram in the bottom panel of Fig.\,\ref{fig:gaia}. We note that 198 nondegenerate companions have multiple radial velocity measurements from both {\em Gaia} DR2 and the other spectroscopic catalogs we considered, while 778 and 116 nondegenerate companions only have radial velocity measurements from either {\em Gaia} DR2 or the other spectroscopic surveys, respectively. 
We also note that 18 isolated white dwarfs from either the SPY or SDSS samples belong to common proper motion pairs. We will comment on the agreement between the radial velocity measurements in Sect.~\ref{sec:analysis-wd} after estimating their gravitational-redshift correction. The {\em Gaia} source identifiers, photometry, and geometric distances of both white dwarfs and nondegenerate stars in common proper motion pairs, along with the element abundances and radial velocities of the latter, are listed in Table\,\ref{tab:cpmp}. The stars with more than one observation in the spectroscopic surveys have repeated entries.

\begin{figure}
    \centering
    \includegraphics{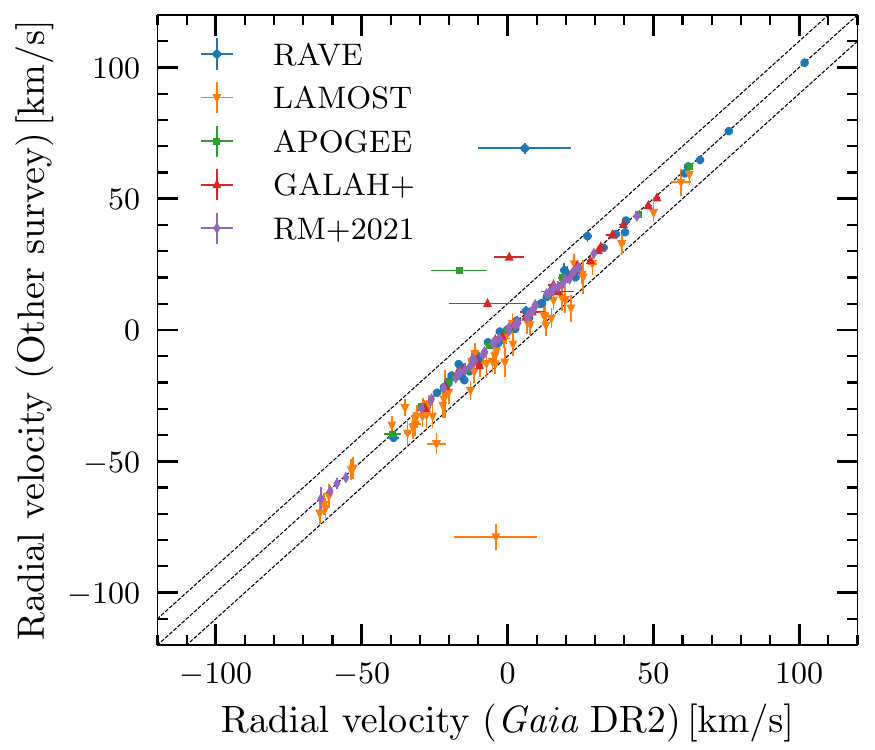}
    \caption{Radial velocity comparison for nondegenerate stars in common proper motion pairs that have measurements in {\em Gaia} DR2 and the other considered surveys. The dashed lines correspond to the equality line and a $\pm 10$\,km\,s$^{-1}$ radial velocity offset.}
    \label{fig:rv_comparison}
\end{figure}

The comparison between the {\em Gaia} DR2 radial velocities and those measured by the other surveys is shown in Fig.\,\ref{fig:rv_comparison}. It confirms an agreement that is well within 10 km\,s$^{-1}$ and supports the use of {\em Gaia} DR2 radial velocities in the kinematic analysis of this sample for the stars that were not observed by other spectroscopic surveys.  The LAMOST radial velocities present a systematic offset of $-4.4 \pm 3.7$\,km\,s$^{-1}$ with respect to the {\em Gaia} DR2 values, as was previously noted \citep{anguiano2018,wang2019}. Therefore we compensate for this offset when we analyze the LAMOST stars in Sect.\,\ref{sec:analysis}. Five data points in Fig.\,\ref{fig:rv_comparison} display a stronger disagreement between \emph{Gaia} DR2 and the other surveys and  have \emph{Gaia} DR2 velocity errors of more than $3$\,km\,s$^{-1}$. We note that the sample of common proper motion pairs that only have {\em Gaia} DR2 radial velocity measurements contains 81 stars with radial velocity errors that are larger than $3$\,km\,s$^{-1}$. We do not know whether these stars  have calibration issues or belong to unresolved binaries. In the remainder of the paper, we will include these objects for statistical purposes, but care should be taken when they are analyzed them  individually.

%--------------------------------------------------------------------
\section{Data analysis}
\label{sec:analysis}
\subsection{Physical parameters of white dwarfs}
\label{sec:analysis-wd}

Because we are interested in studying the relations among the Galactic orbital parameters and stellar ages, we use the white dwarfs as cosmic chronometers. This is possible thanks to the well-constrained physical properties of white dwarfs \citep[][and references therein]{althaus2010}, which enable us to estimate their masses, $M$, and cooling ages, $\tau_{\rm cool}$, that is, the time since a white dwarf formed, as well as their effective temperatures and surface gravities ($T_{\rm eff}$ and $\log{g}$) based on the white dwarf colors and luminosities.  From the white dwarf masses, it is possible to estimate their progenitor masses via the initial-to-final-mass relation (IFMR), and to infer their progenitor ages by means of appropriate stellar evolutionary tracks. 

Our approach follows the idea behind the photometric technique for measuring white dwarf atmospheric parameters, which was first developed by \citet{bergeron1997}. It has since been thoroughly validated and tested against numerous datasets. We used the \emph{Gaia} color-magnitude diagram (Fig.\,\ref{fig:gaia}) to robustly determine the physical parameters of all white dwarfs in our sample through direct interpolation of the observed {\em Gaia} EDR3 magnitudes and colors ($G$, $G_{\rm BP}-G_{\rm RP}$) on the La Plata cooling tracks for He-, CO-, and ONe-core white dwarfs \citep{althaus2013,camisassa2016,camisassa2017,camisassa2019}, which span over a range of $T_{\rm eff} =5000$--80\,000\,K, and $\log{g} = 7$--9.5 or $\log{g} = 7.9$--9.5 for H-rich or H-deficient atmospheres, respectively.  For this purpose, we  computed the absolute magnitudes of the evolutionary models in the {\em Gaia} EDR3 passbands by means of appropriate synthetic spectra for the La Plata cooling sequences with H-rich and H-deficient atmospheres \citep[i.e., using synthetic DA and DB spectra, respectively;][including many unpublished improvements]{koester2010}.
For the interpolation, we used the \verb|scipy| module \citep{scipy} \verb|LinearNDInterpolator|, which is based on the Qhull triangulation\footnote{\url{http://www.qhull.org}}, and we sampled the observed {\em Gaia} EDR3 photometry and parallaxes of each white dwarf with a Monte Carlo method, adopting Gaussian distributions and using the geometric distances of \citet{bailer-jones2021} as priors. We corrected the observed {\em Gaia} EDR3 magnitudes for the effect of interstellar extinction, which we determined by using the geometric distance estimates in combination with the 3D extinction-distance maps of \citet{lallement2019}. We converted the monochromatic extinctions at 5500\,\AA\ of the 3D maps into the {\em Gaia} EDR3 passbands through the \citet{fitzpatrick2019} reddening law with $R_V = 3.1$. 

We adopted cooling tracks of white dwarfs with solar metallicity even when the metallicity of common proper motion pairs was known because the differences in cooling age due to the metallicity content of the white dwarf are typically negligible in contrast to the total age errors or the cooling age differences due to the atmospheric composition (H-rich or H-deficient), which is unknown for most of the white dwarfs in common proper motion pairs. For example, in our sample, a typical white dwarf of 0.6\,M$_{\odot}$ with a DA spectrum will have a cooling age of $\tau_{\rm cool} = 1$\,Gyr at a {\em Gaia} color of $G_{\rm BP} - G_{\rm RP} = 0.35$. In contrast, a white dwarf of the same mass and color, but with a non-DA spectrum, would be 0.4\,Gyr older. This cooling age difference reaches $\sim$1\,Gyr at $G_{\rm BP} - G_{\rm RP} \approx 0.8$, and it reverts back to zero when the white dwarfs have a $G_{\rm BP} - G_{\rm RP} = 1.1$, that is, at $\tau_{\rm cool} \sim 5$\,Gyr. Very few white dwarfs in our sample have cooling ages larger than 5\,Gyr, when the H-rich white dwarfs start to slow down their aging. When a white dwarf was far outside the limits of the evolutionary tracks, as happened for ten isolated white dwarfs and eight members of common proper motion pairs, we did not estimate their physical parameters. Based on the a priori knowledge of white dwarf spectral types, we used the appropriate cooling tracks for H-rich atmospheres (e.g., for the SPY and SDSS samples that have DA spectral types) or H-deficient tracks. As we previously noted, 18 white dwarfs in common proper motion pairs are also part of the SPY or SDSS samples, and another 82 have previous spectral classification in the Montreal White Dwarf Database \citep{dufour2017}. For the majority of white dwarfs in common proper motion pairs, which are currently unclassified, we determined the physical parameters by using both H-rich and -deficient cooling tracks.

\begin{figure}
    \centering
    \includegraphics{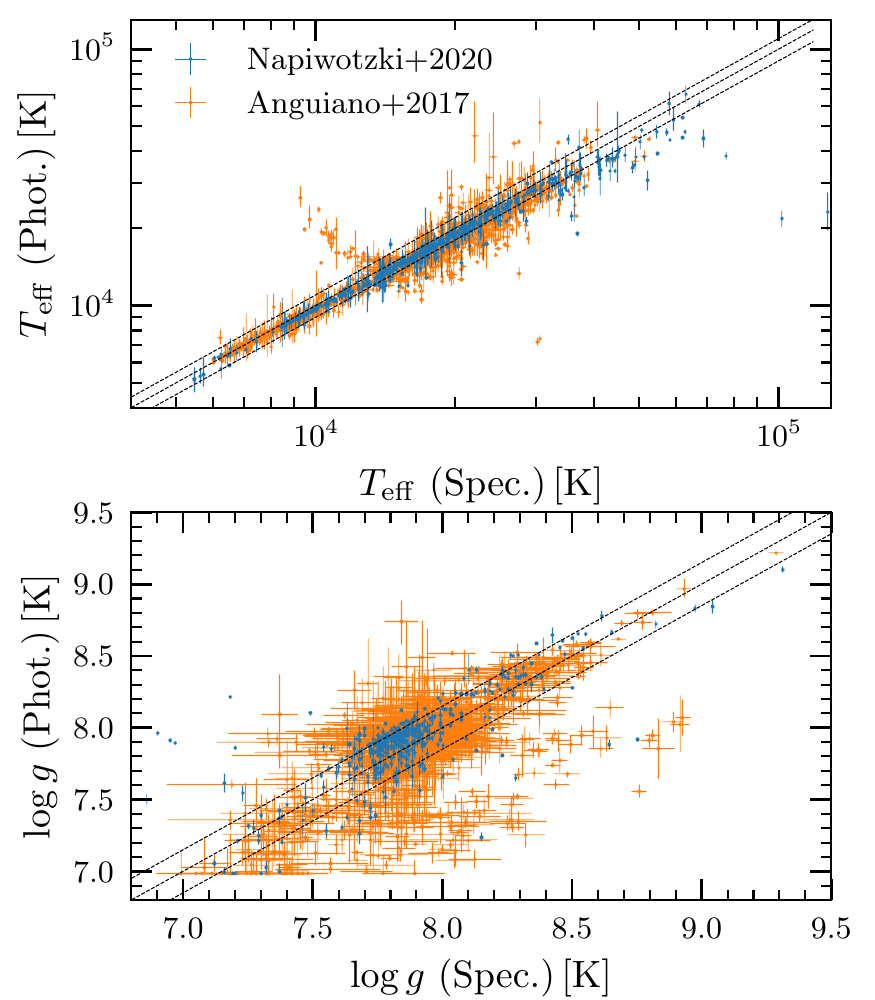}
    \caption{Comparison between spectroscopic and photometric effective temperatures (top panel) and surface gravities (bottom panel) for the white dwarf sample. The dashed lines represent the 10\,\% and 0.15\,dex differences in the $T_{\rm eff}$ and $\log{g}$ comparison plots, respectively.}
    \label{fig:params_comparison}
\end{figure}
\begin{table*}
    \centering
    \setlength{\tabcolsep}{0.11cm}
        \caption{Physical parameters of white dwarfs, interpolated from H-rich evolutionary models.}
        \scriptsize
    \begin{tabular}{@{}P{0.9cm}@{}
    P{1.6cm}@{}  
    P{1.3cm}@{} 
    P{1.3cm}@{} 
    P{0.8cm}@{} 
    D{.}{.}{8}@{} 
    D{.}{.}{8}@{} 
    D{.}{.}{10}@{} 
    D{.}{.}{8}@{} 
    D{.}{.}{10}@{} 
    P{2.4cm}@{}
    c
    D{,}{\,\pm\,}{7}@{}
    p{0.4cm}@{}} 
    \toprule
    \toprule
  System   
  & \multicolumn{1}{c}{$\log{T_{\rm eff}\,[{\rm K}]}$}   
  & $\log{g\,[{\rm cgs}]}$  
  & $M$/M$_{\odot}$ 
  & $R$/R$_{\odot}$ 
  & \multicolumn{1}{c}{$\tau_{\rm cool}$\,[Gyr]}  
  & \multicolumn{1}{c}{$M_i$/M$_{\odot}$ $^{(1)}$}  
  & \multicolumn{1}{c}{$\tau_{\rm prog}$\,[Gyr]$^{(1)}$}  
  & \multicolumn{1}{c}{$M_i$/M$_{\odot}$ $^{(2)}$}  
  & \multicolumn{1}{c}{$\tau_{\rm prog}$\,[Gyr]$^{(2)}$}  
  & Name & SpT  
  & \multicolumn{1}{c}{$\varv_{\rm rad}$\,[km\,s$^{-1}$]$^{(3)}$} 
  & Sample\\
    \midrule 
0001  & $4.066\pm0.015$ & $7.58\pm0.09$ & $0.43\pm0.03$ & 0.018 & 0.540^{+0.085}_{-0.142} &  &  & &  & J030342.23$+$005310.4 & DA & -30.0,16.3 & SDSS\\
0002 & $4.084\pm0.015$ & $7.89\pm0.06$ & $0.54\pm0.03$ & 0.014 & 0.331^{+0.017}_{-0.017} & 1.22^{+0.29}_{-0.22} & 5.837^{+6.544}_{-2.992} & 1.02^{+0.24}_{-0.15}& 11.382^{+9.247}_{-6.155} & J025709.00$+$004628.1 & DA & -17.8,14.4 & SDSS\\
. & &&&&&&&&&&&&\\
3151 & $3.865\pm0.005$ & $7.92\pm0.03$ & $0.55\pm0.02$ & 0.013 & 1.254^{+0.039}_{-0.033} & 1.24^{+0.17}_{-0.15} & 5.539^{+3.623}_{-1.993} & <1.05 & >9.722 & J205905.90$-$003347.8 & DA & -60.2,14.8 & SDSS\\
    \bottomrule
    \multicolumn{14}{l}{{\bf Notes.} The full table is available at the CDS.}   \\
\multicolumn{14}{l}{(1): Progenitor mass determined via the \citet{catalan2008b}
IFMR relation and corresponding progenitor age.}   \\
\multicolumn{14}{l}{(2): Progenitor mass determined via the \citet{cummings2018}
IFMR relation and corresponding progenitor age.}   \\
\multicolumn{14}{l}{(3): $\varv_{\rm rad}$ is the barycentric radial velocity of the white dwarf, which is corrected for gravitational redshift.} \\

    \end{tabular}
    \label{tab:parameters}
\end{table*}

\begin{table*}
\centering
    \setlength{\tabcolsep}{0.101cm}

        \caption{Like Table\,\ref{tab:parameters}, but interpolated from H-deficient evolutionary models.}
        \scriptsize
 \begin{tabular}{@{}P{0.9cm}@{}
    P{1.7cm}@{}  
    P{1.7cm}@{} 
    P{1.7cm}@{} 
    P{1.0cm}@{} 
    D{.}{.}{8}@{} 
    D{.}{.}{8}@{} 
    D{.}{.}{10}@{} 
    D{.}{.}{8}@{} 
    D{.}{.}{10}@{} 
    P{2.2cm}@{}
    c
    p{0.6cm}} 
    \toprule
    \toprule
  System  
  & \multicolumn{1}{c}{$\log{T_{\rm eff}\,[{\rm K}]}$}   
  & $\log{g\,[{\rm cgs}]}$  
  & $M$/M$_{\odot}$ 
  & $R$/R$_{\odot}$ 
  & \multicolumn{1}{c}{$\tau_{\rm cool}$\,[Gyr]}  
  & \multicolumn{1}{c}{$M_i$/M$_{\odot}$ $^{(1)}$}  
  & \multicolumn{1}{c}{$\tau_{\rm prog}$\,[Gyr]$^{(1)}$}  
  & \multicolumn{1}{c}{$M_i$/M$_{\odot}$ $^{(2)}$}  
  & \multicolumn{1}{c}{$\tau_{\rm prog}$\,[Gyr]$^{(2)}$}  
  & Name & SpT  
  & Sample\\
    \midrule
0011  & $3.856\pm0.012$ & $8.05\pm0.07$ & $0.60\pm0.05$ & 0.012 & 2.130^{+0.151}_{-0.144} & 1.74^{+0.47}_{-0.42} & 1.890^{+2.642}_{-0.831} & 1.43^{+0.53}_{-0.37}& 3.402^{+6.513}_{-1.948} &  &  &   CPMP\\
0020  & $3.803\pm0.017$ & $7.99\pm0.11$ & $0.56\pm0.07$ & 0.013 & 2.719^{+0.372}_{-0.228} & 1.39^{+0.70}_{-0.44} & 3.813^{+11.296}_{-2.556} & 1.38^{+0.68}_{-0.41}& 3.935^{+10.068}_{-2.638} &  &  &    GALAH+\\
. & &&&&&&&&&&&\\
3148  & $3.947\pm0.033$ & $8.19\pm0.17$ & $0.69\pm0.11$ & 0.011 & 1.472^{+0.404}_{-0.256} & 2.70^{+0.83}_{-1.24} & 0.614^{+2.556}_{-0.335} & 2.70^{+0.66}_{-1.18}& 0.606^{+2.254}_{-0.285} &  &  &   CPMP\\
     \bottomrule
         \multicolumn{13}{l}{{\bf Notes.} The full table is available at the CDS.}   \\
\multicolumn{13}{l}{(1): Progenitor mass determined via the \citet{catalan2008b}
IFMR relation and corresponding progenitor age.}   \\
\multicolumn{13}{l}{(2): Progenitor mass determined via the \citet{cummings2018}
IFMR relation and corresponding progenitor age.}   \\
    \end{tabular}
    \label{tab:parameters_db}
\end{table*}

Because the photometric interpolation procedure also allows us to derive $T_{\rm eff}$ and $\log{g}$,  we compared them with the corresponding spectroscopic values for the SPY and SDSS white dwarfs in Fig.\,\ref{fig:params_comparison} as a way of testing the accuracy of our method.  The SPY sample shows a generally good agreement that is mostly within 10\,\% over the whole $T_{\rm eff}$ range and 0.15\,dex for $\log{g}$, although some larger systematic differences are seen for the hottest or the coolest objects with low $\log{g}$. These differences may arise from the difficulty of obtaining reliable photometric fits for hot white dwarfs on the one hand, but on the other hand, they may also indicate that these white dwarfs could be unresolved binaries or have problematic data. Caution should therefore be taken when studying them further. The SDSS white dwarfs display a much larger scatter that is likely due to their lower spectral resolution, which affects the measurement of spectroscopic parameters. The data shown in Fig.\,\ref{fig:params_comparison} can be visually compared with the in-depth analyses performed by \citet{tremblay2019b} and \citet{bergeron2019}, which showed a similar degree of scatter for the SDSS samples. The comparisons among spectroscopic and photometric measurements of white dwarf masses and cooling ages for the SPY and SDSS samples show average scatter around the zero of $\approx0.1$\,M$_{\odot}$ and $\approx0.3$\,Gyr, respectively. The scatter of cooling ages is temperature dependent, increasing towards low $T_{\rm eff}$. The systematic differences among spectroscopic and photometric determinations, which have been analyzed in particular with regards to the mass distribution of the white dwarf population \citep{tremblay2019b,bergeron2019}, are smaller than other dominant sources of error that in turn affect the determination of white dwarf total ages, as we discuss in the following section. By employing the photometric technique to measure the physical parameters of both the isolated white dwarfs and those in common proper motion pairs, we aim to obtain uniform results and to mitigate possible systematic discrepancies affecting the sample, although we pay the price of larger uncertainties in some cases.

The radial velocities of isolated white dwarfs in the SPY and SDSS samples are  subject to  gravitational redshift. We therefore determined the correction as  $\varv_{\rm g} = {\rm G} M/cR$, where ${\rm G}$ is the gravitational constant, $c$ is the speed of light, and $M/R$ is the mass-radius ratio as determined via the photometric interpolation. Hence, the apparent white dwarf velocities can be corrected to obtain their radial velocity $\varv_{\rm rad} = \varv_{\rm app} - \varv_{\rm g}$. We verified the radial velocity agreement for the 18 common proper motion pairs that have radial velocities for the white dwarfs, either from SPY or SDSS in Table\,\ref{tab:parameters}, and for the nondegenerate companions in Table\,\ref{tab:cpmp}. The six pairs containing SPY white dwarfs have $\varv_{\rm rad} ({\rm WD}) - \varv_{\rm rad} ({\rm CPMP}) = -0.5 \pm 3.8$\,km\,s$^{-1}$, while the 12 pairs containing SDSS white dwarfs have that $\varv_{\rm rad} ({\rm WD}) -  \varv_{\rm rad} ({\rm CPMP}) = -4.8 \pm 3.8$\,km\,s$^{-1}$. We note that these radial velocity differences are on the same order of magnitude as those measured for tangential velocities implied by the selection criteria of Sect.\,\ref{sec:sample_cpmp}.

In total, we were able to determine physical parameters for 2067 out of 2077 isolated white dwarfs. In addition, 25 of these white dwarfs are at the edge of our model grid, causing their parameters to be incompletely determined. We determined physical parameters for 1084 out of 1092 white dwarfs in common proper motion pairs, three of which are at the edge of the H-rich grid and have less accurate parameters. We have determined physical parameters from the H-deficient model grid for only 691 white dwarfs in common proper motion pairs.  The white dwarf physical parameters that were determined from H-rich models are listed in Table\,\ref{tab:parameters}, while those determined from the H-deficient models are listed in Table\,\ref{tab:parameters_db}.  White dwarfs for which we could not retrieve a spectral type are featured in both tables. 
\subsection{White dwarf total ages}
\label{sec:wd_progenitors}

 In order to estimate the white dwarf total ages, we first inferred their progenitor masses ($M_{i}$) using two semi-empirical IFMRs by \citet{catalan2008b} and \citet{cummings2018}. These relations are two- and three-piece linear fits to observations of white dwarfs in open clusters, respectively, and are a good representation of the predominantly solar-metallicity Galactic disk population analyzed in this paper \citep[see e.g.][]{romero2015}. The adopted IFMRs imply slightly different progenitor masses, with lower limits for progenitor masses of of 1\,M$_{\odot}$ \citep[][]{catalan2008b} and 0.83\,M$_{\odot}$ \citep{cummings2018} for white dwarf masses of 0.52 and 0.55\,M$_{\odot}$, respectively. We extrapolated both relations down to progenitor masses $M_{i} =0.8$\,M$_{\odot}$, which would be as old as the Milky Way age ($\sim$13.5\,Gyr).
 
 The progenitor ages ($\tau_{\rm prog}$) of white dwarfs were estimated from the BaSTI evolutionary tracks of appropriate progenitor mass, [Fe/H], and [$\alpha$/Fe] \citep{hidalgo2018,pietrinferni2021}. We adopted the composition of the nondegenerate companions for the white dwarfs in common proper motion pairs (Table\,\ref{tab:cpmp}). The adopted error on RAVE abundances is 0.2\,dex \citep{kunder2017}. We also adopted an additional 0.05\,dex abundance uncertainty that we sum in quadrature with the listed errors for LAMOST \citep{xiang2019},  APOGEE \citep{jonsson2020}, and GALAH+ \citep{buder2021}.  When [Fe/H] and [$\alpha$/Fe] lay beyond the grid limits, we assumed the corresponding nearest values. When the metallicity of a white dwarf is unknown (e.g., that of the isolated white dwarfs), we assumed a Gaussian distribution of [Fe/H]~$ = 0.00 \pm 0.26$\,dex and [$\alpha$/Fe]~$=0$\,dex, as is typical for our sample (see Section\,\ref{sec:discussion}). If the nondegenerate companions of white dwarfs in common proper motion pairs were observed by two different spectroscopic surveys, we determined their progenitor parameters for both the measured abundances. While these values typically agree within the errors,  we note that the progenitor lifetimes could differ by a few million years to up to billion years, depending on the adopted parameters. 

\begin{figure*}
    \centering
    \includegraphics{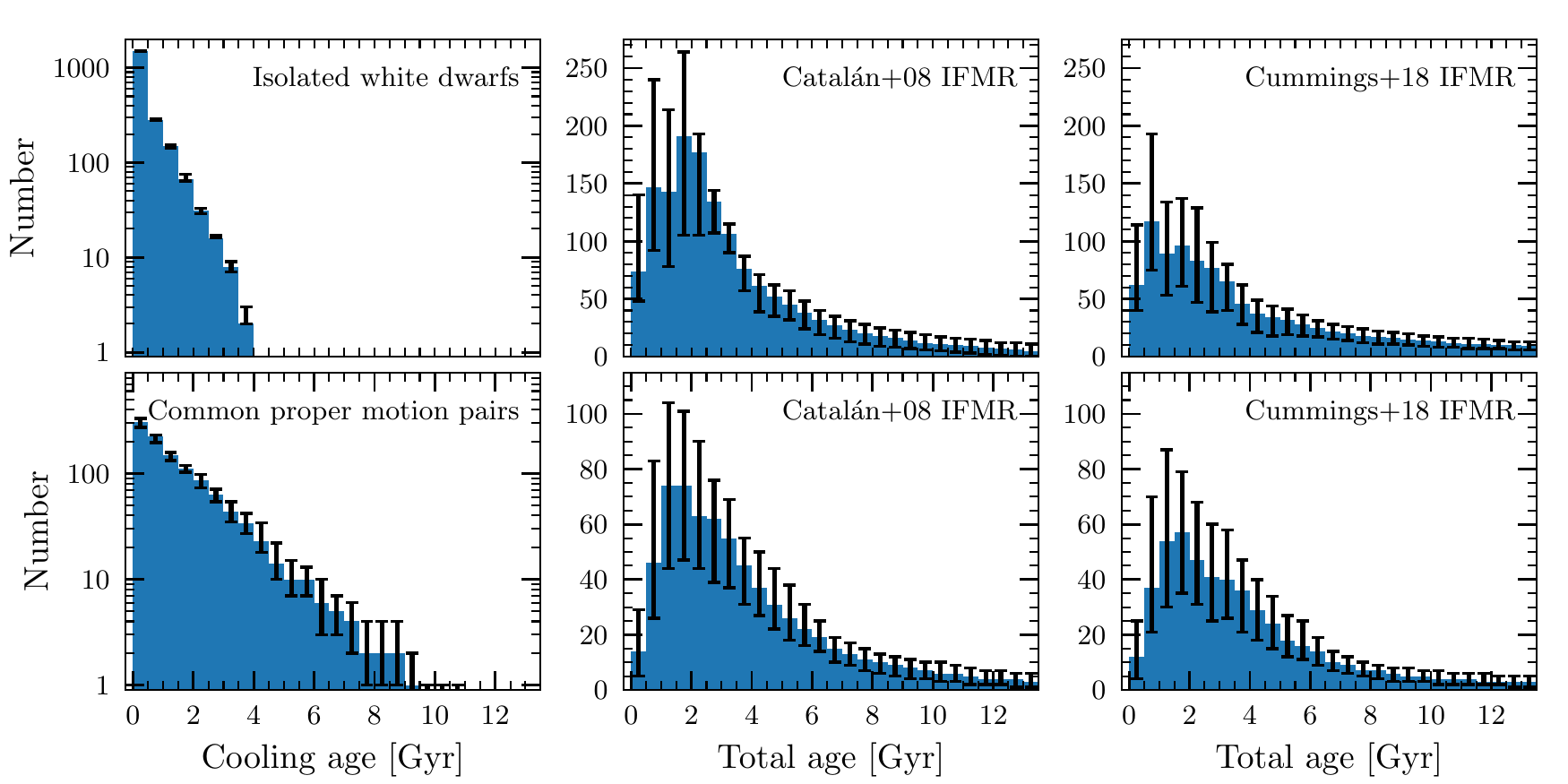}
    \caption{Cooling age and total age ($=\tau_{\rm cool} + \tau_{\rm prog}$) distributions for the white dwarf samples (top panels) and the white dwarfs in common proper motion pairs (bottom panels). The total ages are shown for both our adopted IFMRs. The histogram bars and the error bars account for a standard 2/8 ratio of non-DA/DA spectral types and for the age uncertainties, as they are listed in Tables\,\ref{tab:parameters} and \ref{tab:parameters_db}.}
    \label{fig:hist_ages_single}
\end{figure*}

 The distributions of cooling ages and total ages for the considered samples are plotted in Fig.\,\ref{fig:hist_ages_single}, where we display the median age bin occupancy and the 16--84\% scatter, which we determined through a Monte Carlo sampling of the uncertainties. The histogram bars of common proper motion pairs also take a 2/8 ratio of non-DA/DA white dwarfs into account when the spectral type is undetermined. Isolated white dwarfs   have typically younger cooling ages because they were selected from magnitude limited spectroscopic samples, which favor intrinsically bright/young objects. The common proper motion sample is less affected by such a selection bias. However, we note that as expected for a predominantly disk population, both samples have similar total ages distributions. Because of the strong correlation between stellar mass and age, especially in the low mass regime, the progenitor ages corresponding to initial masses computed via the \citet{cummings2018} IFMR are typically longer than those computed via the \citet{catalan2008b} formulation. White dwarfs of $\approx$0.5\,M$_{\odot}$ have very uncertain total ages because their progenitors of $\approx$0.8\,M$_{\odot}$ have long evolutionary time scales and the relation between initial mass and age is very steep for such low-masses. Moreover, because the evolutionary timescales of $\approx 0.8$\,M$_{\odot}$ are close to or even longer than the age of the Milky Way ($\sim$13.5\,Gyr),  we were often only able to derive upper limits. Hence, the dominant source of scatter in our age determinations originates from the error propagation through the IFMRs, followed by those implied by [Fe/H] and [$\alpha$/Fe] uncertainties. 
 The estimated progenitor masses and ages are listed in  Table\,\ref{tab:parameters} and \,\ref{tab:parameters_db}, respectively, for H-rich and H-deficient cooling models.
\subsection{Galactic orbit integration}
\label{sec:analysis-kin}
We used the {\sc python} package for galactic dynamics, \verb|galpy|\footnote{\url{http://github.com/jobovy/galpy}} \citep[v1.6;][]{bovy2015}, to compute the Galactic orbital parameters of the white dwarfs. In the \verb|galpy| framework, we adopted the standard Galactic potential, \verb|MWPotential2014|, that is defined as a combination of a spherical bulge, plus disk \citep[][]{miyamoto1975} and halo components \citep{navarro1997}, which we scaled to the galactocentric distance of the Sun, $R_0 = 8.178$\,kpc \citep{gravity2019} and adopted the rotation velocity at the solar circle, $\Theta_{\rm 0} = 235$\,km\,s$^{-1}$ \citep[][]{reid2004,gravity2019}. The \citet{schoenrich2010} results for the solar motion components were also adopted. We sampled the observed data using a Monte Carlo method (coordinates, proper motions, radial velocities, and geometric distances), and we randomly drew 10\,000 combinations of initial parameters for each star, assuming Gaussian distributions and accounting for the {\em Gaia} EDR3 correlation matrix. In this way, we computed their Cartesian coordinates ($X$, $Y$, $Z$) and the corresponding velocity components ($U$, $V$, $W$)  in a left-handed galactocentric rest frame, that is, with the $x$-axis pointing from the Galactic center towards the Sun and $y$-axis along the direction of the Galactic rotation. For ease of computing time, we used the fast orbit-characterization functionality of \verb|galpy| \citep{mackereth2018} that enables to estimate, as a first guess, the Galactic orbital periods, via the St\"ackel approximation \citep{binney2012}. Subsequently, using their randomly sampled initial conditions, we integrated the Galactic orbits of each star for ten cycles around the Milky Way that we sampled at fifty time intervals. After the integration, we numerically estimated the Galactic orbital parameters such as eccentricity, azimuthal action (i.e., the vertical component of the angular momentum in the considered axisymmetric potential), Galactic pericenter and apocenter, maximum height above the Galactic plane ($e$, $L_z$, $R_{\rm peri}$, $R_{\rm apo}$, $Z_{\rm max}$, respectively). The Galactic Cartesian coordinates and their velocity components, as well as the Galactic orbital parameters for a subset of the studied sample are listed in Table\,\ref{tab:kinematics}, while the full table is available at the CDS.

\section{Results and discussion}
\label{sec:discussion}
\subsection{Spatial distribution and surveyed volume} 
\begin{table*}
    \centering
        \setlength{\tabcolsep}{0.101cm}
\renewcommand{\arraystretch}{1.2}
        \caption{Galactic Cartesian coordinates, Cartesian velocity components in the local standard of rest, and Galactic orbital parameters of isolated white dwarfs and common proper motion pairs. }
        \scriptsize
    \begin{tabular}{@{}P{1.0cm}@{}
   P{1.5cm}@{}
    c 
    c 
    c 
    D{,}{\,\pm\,}{5} 
    D{,}{\,\pm\,}{5}
    D{,}{\,\pm\,}{5}
    D{.}{.}{6}@{} 
    D{,}{}{4}@{} 
    D{.}{.}{6}@{} 
    D{.}{.}{6}@{} 
    D{.}{.}{6}@{}} 
    \toprule
    \toprule
  System 
  & Sample 
  & \multicolumn{1}{c}{$X$\,[kpc]} 
  & \multicolumn{1}{c}{$Y$\,[kpc]} 
  & \multicolumn{1}{c}{$Z$\,[kpc]} 
  & \multicolumn{1}{c}{$U$\,[km\,s$^{-1}$]} 
  & \multicolumn{1}{c}{$V$\,[km\,s$^{-1}$]} 
  & \multicolumn{1}{c}{$W$\,[km\,s$^{-1}$]} 
  & \multicolumn{1}{c}{$e$} 
  & \multicolumn{1}{c}{$L_z$\,[kpc\,km\,s$^{-1}$]} 
  & \multicolumn{1}{c}{$R_{\rm peri}$\,[kpc]} 
  & \multicolumn{1}{c}{$R_{\rm apo}$\,[kpc]} 
  & \multicolumn{1}{c}{$Z_{\rm max}$\,[kpc]} \\
    \midrule
0001 & SDSS & $8.329$ &  $+0.008$ &  $-0.144$ &  +63.95,10.97 &  +11.47,0.76 &  -0.28,12.03 & 0.21^{+0.03}_{-0.03} & 2053,^{+5}_{-5} & 7.25^{+0.23}_{-0.21} & 11.11^{+0.45}_{-0.37} & 0.21^{+0.13}_{-0.03} \\ 
0002 & SDSS & $8.279$ &  $+0.008$ &  $-0.094$ &  +37.05,9.42 &  +10.02,0.90 &  +8.16,10.80 & 0.13^{+0.03}_{-0.02} & 2029,^{+6}_{-6} & 7.69^{+0.20}_{-0.20} & 9.94^{+0.32}_{-0.24} & 0.17^{+0.16}_{-0.05} \\ 

  .       &  \\
3151 & SDSS & $8.134$ &  $+0.049$ &  $-0.015$ &  -53.32,8.67 &  -35.33,9.70 &  -11.37,7.03 & 0.23^{+0.05}_{-0.05} & 1622,^{+77}_{-80} & 5.47^{+0.53}_{-0.51} & 8.83^{+0.07}_{-0.06} & 0.16^{+0.10}_{-0.09} \\ 
         \bottomrule
         \multicolumn{13}{l}{{\bf Notes.} The full table is available at the CDS.}  
    \end{tabular}
    \label{tab:kinematics}
\end{table*}
\begin{figure*}
    \centering
    \includegraphics[width=0.9\linewidth]{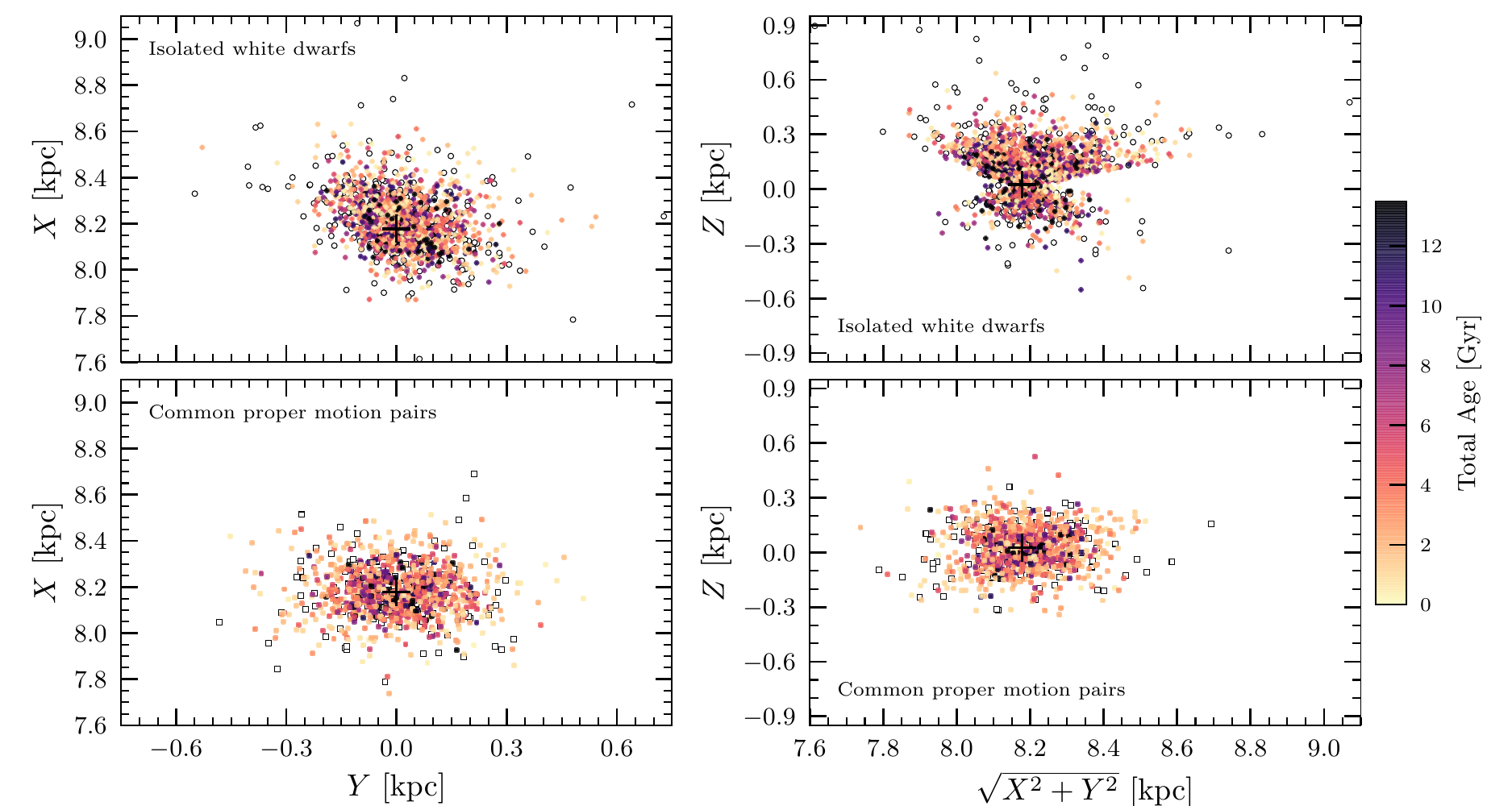}
    \caption{Spatial distribution of single white dwarfs (top panels) and common proper motion pairs (bottom panels) in the galactocentric rest frame. Each point is color-coded according to the median total age of the represented star, which is mapped in the color bar on the right. The Sun's location at (X, Y, Z) = (8.178, 0, 0.025)\,kpc is marked by a cross. White dwarfs with $M + \sigma_M \leq 0.52$\,M$_{\odot}$ are plotted as white symbols.}
    \label{fig:xyz}
\end{figure*}
The two samples we studied have a median geometric distance of $\approx 160$\,pc, and spatial distributions with long tails stretching up to 0.5 and 1\,kpc away from the Sun for the isolated white dwarfs and the common proper motion pairs, respectively. Due to our strict selection cuts and quality requirements that are presented in Sect.\,\ref{sec:sample}, just four common proper motion pairs and 15 isolated white dwarfs are found within 20\,pc from the Sun, while 17 common proper motion pairs and 83 isolated white dwarfs are at less than 40\,pc away from the Sun \citep[cf. 144 and 1263 systems, which lay within 20 and 40\,pc from the Sun, respectively, as identified with {\em Gaia} DR2 and EDR3 by][]{hollands2018, tremblay2020, mccleery2020, gentilefusillo2021}. Similarly, just 561 isolated white dwarfs and 213 common proper motion pairs are within 100\,pc from the Sun \citep[cf 13,732 white dwarfs in {\em Gaia} DR2;][]{jimenez-esteban2018,torres2019}.  

The spatial distribution of the studied stars is shown in Fig.\,\ref{fig:xyz}, where we plot the isolated white dwarfs in the top panels and the common proper motion pairs in the bottom panels, color-coding each point according to their estimated total ages.  The two samples are roughly symmetrically distributed around the Sun in the galactocentric $X$ and $Y$ coordinates, with a dispersion of 100--130\,pc.  In the vertical direction with respect to the Galactic plane, the SDSS white dwarfs extend farther out in the northern hemisphere at an average $Z = 120$\,pc and with a dispersion of 150\,pc, while the SPY white dwarfs and the common proper motion pairs are more symmetrically distributed around $Z = 0$\,pc with a dispersion of 100\,pc. No strong correlations among the spatial coordinates and ages are observed, but  some sort of clustering is visible that could be investigated in future work.  White dwarfs of $M+\sigma_M \leq 0.52$\,M$_{\odot}$, which are not color-coded in Fig.\,\ref{fig:xyz} because their total ages are typically more uncertain, appear slightly more scattered than higher mass white dwarfs, displaying a dispersion of a few 10\,pc larger than that of the whole sample. This larger spread is likely due to their intrinsically brighter magnitudes that enabled them to be detected at larger distances. We note that about 5\% of the studied white dwarfs also have masses below 0.45\,M$_{\odot}$ at a 3-$\sigma$ level, which may indicate that their evolution was affected by yet unresolved companions or by their common proper motion companions.

\subsection{Kinematic and dynamic properties} 
\begin{figure*}
    \centering
    \includegraphics[width=0.9\linewidth]{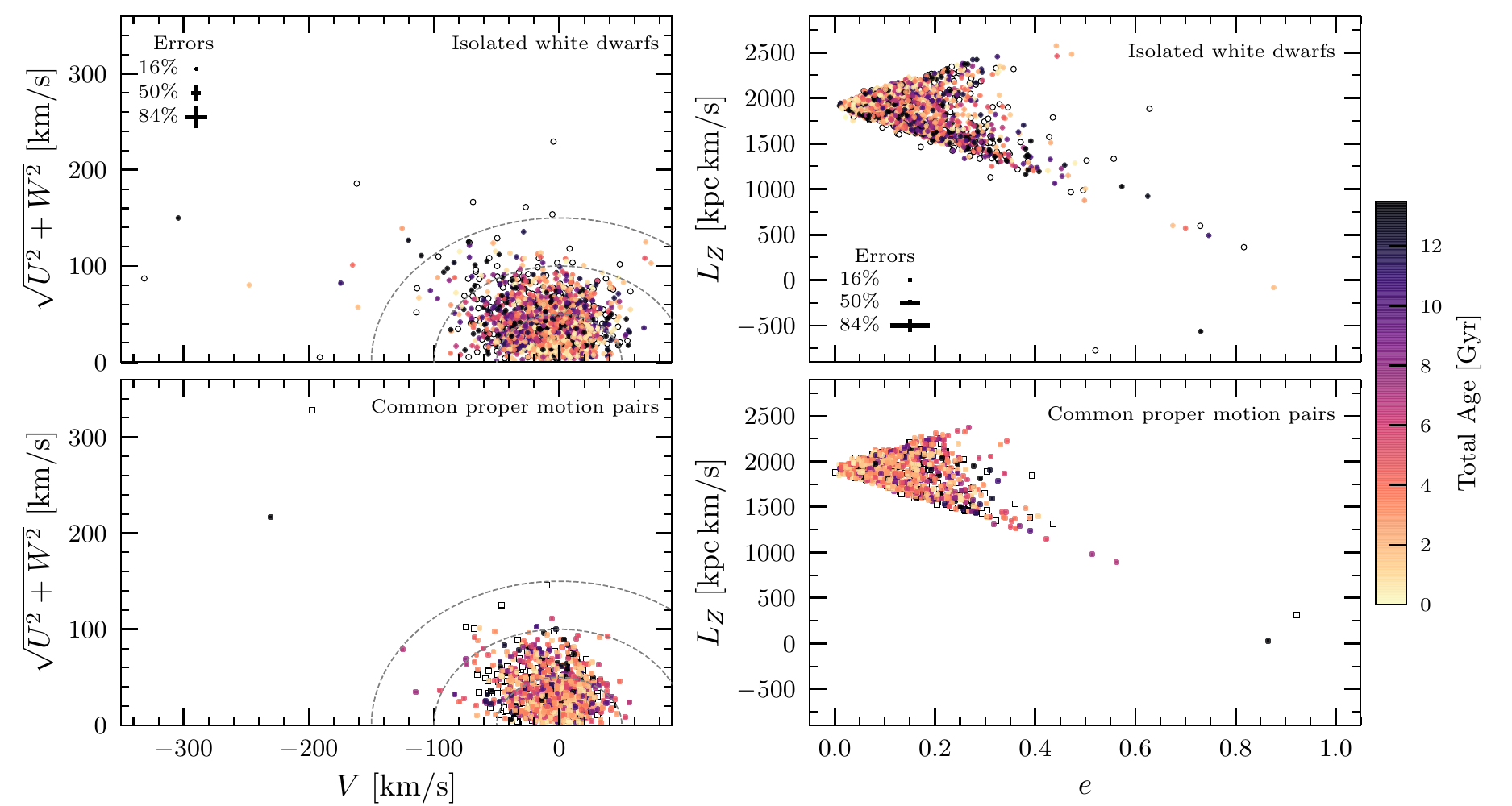}
    \caption{Toomre diagram (left panels) and azimuthal action vs. eccentricity (right panels) of single white dwarfs (top panels) and common proper motion pairs (bottom panels). The velocity components are in the rotating rest frame. Colors and symbols are the same as in Fig.\,\ref{fig:xyz}. The quantiles of the error bar distributions are also plotted for the isolated white dwarfs. The dashed curves in the Toomre diagrams (left panels) are circles with diameters of 100, 200, 300 km\,s$^{-1}$.}
    \label{fig:kinematics}
\end{figure*}

The kinematics of single white dwarfs in the SPY and SDSS samples was previously studied by \citet{pauli2003,pauli2006}, \citet{richter2007}, \citet{dimpel2018}, and \citet{anguiano2017}, respectively, and our results do not drastically change those already obtained. Nevertheless, the higher accuracy and precision of the {\em Gaia} EDR3 astrometry significantly reduces the uncertainties on the Galactic orbital parameters. Especially the accuracy for the SDSS sample should have been increased for two reasons: (i) the {\em Gaia} parallaxes reduced the scatter of atmospheric parameters, leading to improved estimates of the white dwarf masses, hence gravitational redshifts, and cooling ages, and (ii) we calibrated the radial velocities of the SDSS white dwarfs against those of the high-resolution measurements for the SPY sample. 

In Fig.\,\ref{fig:kinematics}, we show two diagrams that are commonly used to study the stellar kinematics: on the left, the Toomre diagram, which uses the Galactocentric Cartesian velocity components, and on the right, the vertical component of the angular momentum versus the eccentricity of Galactic orbits. The single white dwarfs and the common proper motion pairs are displayed in the top and bottom panels, respectively. In general, as for the spatial distribution, there is no strong correlation among velocity components and total ages, as would be expected for a mixed population. We note a larger scatter in the parameter spaces for isolated white dwarfs with respect to common proper motion pairs. The majority of systems belong to a thin disk population, with velocities within $\pm 100$\,km\,s$^{-1}$ of the general motion around the Solar circle in the Toomre diagrams. These stars also mostly have quasi-circular Galactic orbits with $e \lesssim 0.3$ in the two right-hand side panels of Fig.\,\ref{fig:kinematics}. Moreover, we also note an unusually high number of isolated white dwarfs of typically old ages ($\geq10$\,Gyr) that have small velocity components  ($\sqrt{U^2+V^2+W^2}<100$\,km\,s$^{-1}$) and orbits with low eccentricity ($e<0.2$). 

Learning from the kinematic classification scheme drawn by \citet{pauli2006}, we more simply considered stars with $e<0.27$ as the most likely thin-disk members, while those with $0.27 \leq e \leq 0.7$ as thick-disk members. The candidate halo stars either have the most eccentric Galactic orbits with $e>0.7$ and $L_Z >0$\,kpc\,km\,s$^{-1}$ or are retrograde orbits with $L_Z < 0$\,kpc\,km\,s$^{-1}$. The isolated white dwarf sample accounts for 90\,\% and 10\,\% of thin- and thick-disk members, respectively. The common proper motion pairs are instead divided into 95\,\% and 5\,\% of thin- and thick-disk members, respectively. There are only seven isolated white dwarfs and two common proper motion pairs that are halo candidates, corresponding to less than 0.3\,\% of the respective samples. While this result is just a factor of three smaller than the halo fraction identified by \citet{torres2019} in their analysis of the 100-pc {\em Gaia} DR2 white dwarf sample, pre-{\em Gaia} results by \citet{pauli2006}  and \citet{richter2007} found a 2\,\%  and $<5$\,\%, respectively, of halo members in the SPY sample. A {\em Gaia} DR2 update on this sample, leveled down the estimate to around 1\,\% \citep[][]{dimpel2018}. It is important to precisely establish the fraction of halo white dwarfs because they have long been debated as important contributors to the baryonic dark matter. While the SPY and SDSS samples include relatively warm, younger white dwarfs with Balmer absorption lines that could belong to a smaller population of halo white dwarfs, previous works, although with contrasting results, focused on the kinematic analysis of much cooler white dwarfs that do not possess suitable absorption lines to measure radial velocities \citep{oppenheimer2001,reid2001}. The inclusion of common proper motion pairs, as we do here, may help to solve this issue in future because the nondegenerate companions enable to measure the systemic radial velocities.  

The three kinematic sub-groups display a wide range of ages. While those of thin-disk stars strongly peak at young ages, mimicking the distributions displayed in Fig.\,\ref{fig:hist_ages_single}, thick-disk and halo candidates have more scattered age distributions. Within the group of halo candidates, four out of seven isolated white dwarfs have lower masses  ($M<0.55$\,M$_{\odot}$), for which the total ages are rather uncertain and typically very large, as expected for such an old population. The isolated white dwarfs that were identified as halo candidates are numbers 0703, 1461, 1463, 1536, 1762, 2014, 2921, which are known as WD 1448$+$077, WD\,2351$-$368, WD\,2359$-$324, HE\,0201$-$0513, SDSS\,J225513.66$+$230944.1, WD\,1314$-$153, and WD\,1524$-$749, respectively. All the SPY white dwarfs except for WD\,1314$-$153 were previously classified by \citet{pauli2006} as thick-disk or halo members; WD\,1314$-$153 is a new addiction to the \citet{pauli2006} sample, but it was also classified as a halo member by \citet{richter2007} and confirmed by \citet{dimpel2018}. The white dwarf HS\,1527+0614, which was classified by \citet{pauli2006} as another halo member,  was re-classified as a thin disk object \citep[confirming previous results based on {\em Gaia} DR2;][]{dimpel2018}. Another SPY white dwarf, WD\,0252--350 which is a suggested halo member \citep{pauli2006,dimpel2018}, did not pas our quality cuts. The SDSS white dwarf J225513.66$+$230944.1 is a new identification. The two common proper motion pairs that we identified as halo candidates are numbers 0776 and 2561, in which the white dwarfs that do not yet have a spectroscopic classification are SDSS\,J151530.71$+$191130.8 and LSPM\,J1756$+$0931S, respectively. The two white dwarfs have relatively low masses of $M\leq 0.5$\,M$_{\odot}$. These two systems only have {\em Gaia} DR2 radial velocity measurements for their nondegenerate members, and that of number 2561, LSPM\,J1756+0931N, has the largest measurement in our sample, that is $\varv_{\rm rad} = -324.9 \pm 0.7$\,km\,s$^{-1}$. This value is supported by a relatively large proper motion of 225\,mas\,yr$^{-1}$ that corresponds to 77\,km\,s$^{-1}$ at a distance of 72\,pc. 

The random forest classification of the 100-pc sample presented by \citet{torres2019}  used {\em Gaia} DR2 astrometry and ages, but did not include radial velocities. Three of their halo candidates are isolated white dwarfs in our sample (\#\,0703, 1461, and 2014), which were also classified as halo members above. The other 682 stars in common with the \citet{torres2019} sample were classified by these authors as thin- and thick-disk members, which also confirms our results, although with a large overlap between the two samples. In general, we note that future classifications of thin/thick disk and halo members will greatly benefit from precise and accurate radial velocity measurements.

The Galactic orbits of the stars we studied reach a wide range of apocenters and pericenters that mostly span between $R_{\rm peri} = 5$--7\,kpc, that is, within the solar circle, and $R_{\rm apo} = 9$--10\,kpc. On the other hand, the high-velocity outliers (ten isolated white dwarfs and four common proper motion pairs with $V < -100$\,km\,s$^{-1}$ that also have $e > 0.5$ and are likely thick-disk or halo members), reach down to $R_{\rm peri} \lesssim 3$\,kpc, but have Galactic orbits mostly constrained within $R_{\rm apo} = 8$--12\,kpc. Moreover, the common proper motion pair 0776, which hosts a white dwarf of $\approx 0.34$\,M$_\odot$, has the most extreme trajectory spanning between $R_{\rm peri} \approx 1$\,kpc and $R_{\rm apo} \approx 25$\,kpc. The fact that these 14 systems have $R_{\rm peri} \lesssim 3$\,kpc may suggest their dynamical interaction with the Galactic bar and/or bulge potentials. The isolated white dwarf 1762 is the record holder with  $R_{\rm peri} \approx 0.5$\,kpc. Furthermore, there is a group of four isolated white dwarfs (numbers 0038, 0855, 0885, and 0886) that also reach large $R_{\rm apo} = 19$--22\,kpc although they reach $R_{\rm peri} =5$--8\,kpc and have Galactic thick-disk orbits of moderate eccentricity.  The most interesting object in this group is  number 0886, which is an ultra-massive DA white dwarf, LP\,387--21, which has $M = 1.24$\,$M_{\odot}$. The total age of this white dwarf is dominated by its cooling age, $\tau_{\rm cool} \approx 2$\,Gyr, because the predicted progenitor mass is in excess of 6.5\,M$_{\odot}$. Nevertheless, this white dwarf could also be a binary-merger product, in which case its progenitor life time and cooling age would be much longer \citep{wegg2012,cheng2020, temmink2020}. Moreover, we note that 0886 is located in the ``Q branch'', that is, the region of the cooling sequence in which core crystallization takes place in massive white dwarfs \citep[][]{tremblay2019}. In this case as well, the cooling time may therefore be longer than what is currently estimated due to the release of latent heat during Ne$^{22}$ settling in ultra-massive white dwarfs \citep{bauer2020,blouin2021,camisassa2021}. A longer total age might therefore push this star towards a halo membership.

The maximum vertical displacements reached by the studied sample of white dwarfs are typically within $Z_{\rm max} < 2$\,kpc, but 26 isolated white dwarfs and two common proper motion pair stars that reach $Z_{\rm max} > 2$\,kpc have $|W|>60$\,km\,s$^{-1}$. The white dwarfs in this group are mostly thin- and thick-disk members, but there are also two halo candidates (numbers 1463 and 2561).

\begin{figure}
    \centering
    \includegraphics[width=0.9\linewidth]{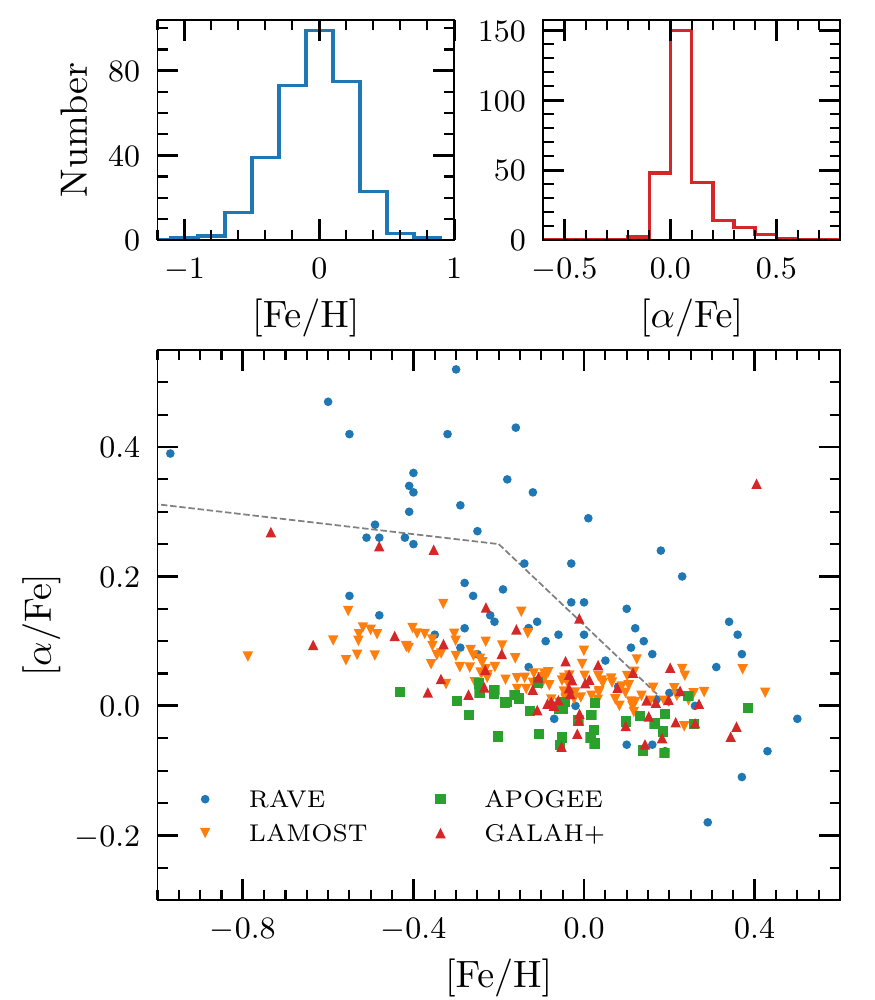}
    \caption{Histogram distributions of [Fe/H] and [$\alpha$/Fe] (top panels) and their correlation (bottom panel) for the sample. The dashed line is the reference curve from \citet{hayden2015}, which reflects the typical trend of disk stars. }
    \label{fig:chemistry}
\end{figure}
\subsection{Common proper motion pairs with abundances}
\begin{figure*}
    \centering
    \includegraphics[width=0.9\linewidth]{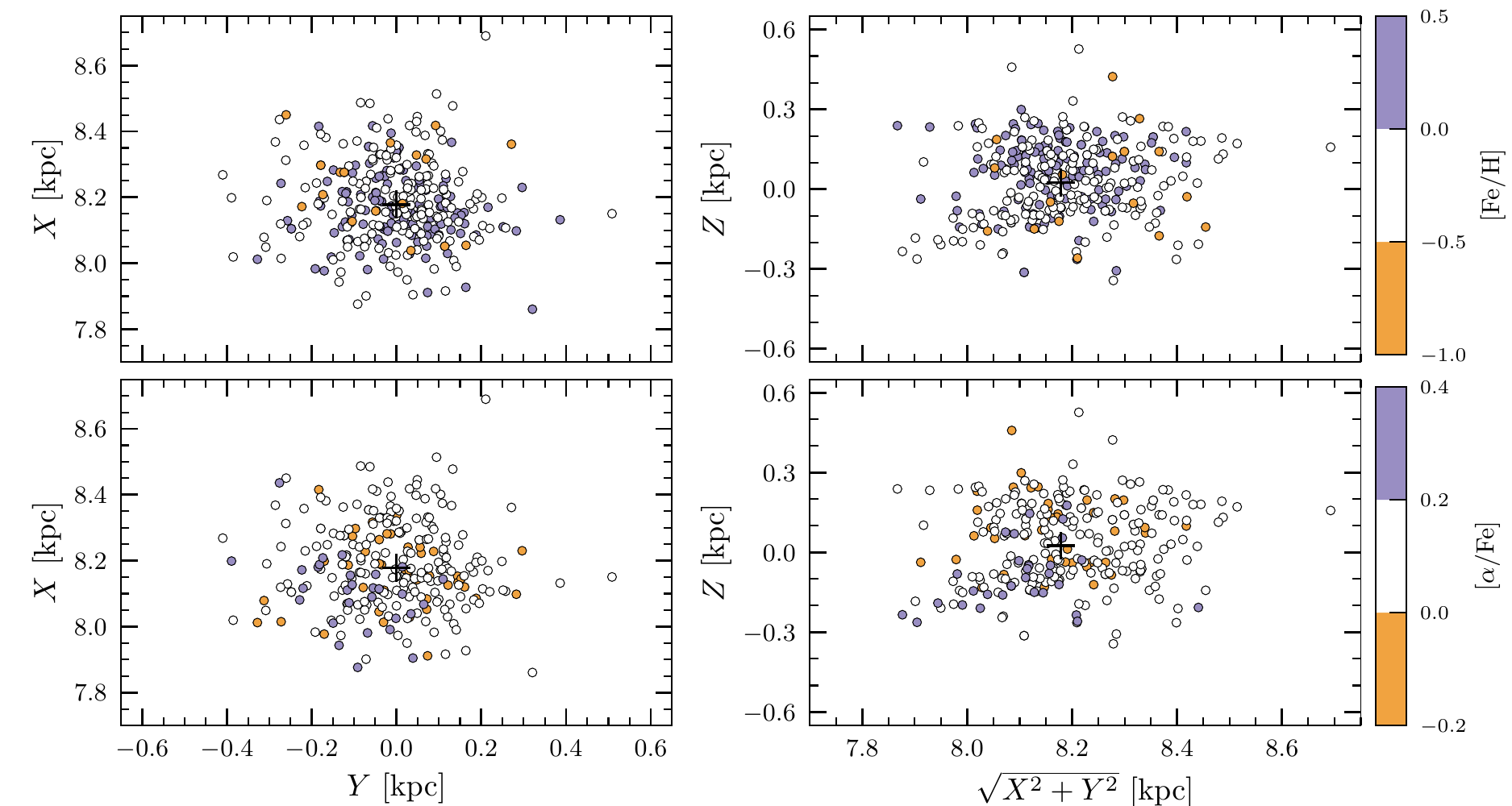}
    \caption{Spatial distribution of the common proper motion pairs that we have color-coded according to the measured [Fe/H] and [$\alpha$/Fe] of the nondegenerate stars in the top and bottom panels, respectively. The adopted color scheme is shown on the right.}
    \label{fig:xyz_feh-alpha}
\end{figure*}

In the top panels of Fig.\,\ref{fig:chemistry}, we show the histogram distributions of [Fe/H] and [$\alpha$/Fe] of the 314 common proper motion pairs that have available spectra for the nondegenerate stars. They have [Fe/H]~$= 0.00 \pm 0.26$\,dex and predominantly [$\alpha$/Fe]~$>0$\,dex. The correlation between [Fe/H] and [$\alpha$/Fe] that is seen in the bottom panel of Fig.\,\ref{fig:chemistry} is typical for thin-disk stars in the Solar neighborhood \citep[cf][]{navarro2011,adibekyan2011,hayden2015}, but some interlopers from the thick disk may be present. Moreover, we note that while the stars with LAMOST, APOGEE, and GALAH+ spectra  mostly have [$\alpha$/Fe]~$ < 0.2$\,dex, which is more common for local thin disk stars, the RAVE spectra have [$\alpha$/Fe]~$ > 0.2$ that is more common for thick disk objects. The iron and $\alpha$-element abundances of stars that have more than one spectra roughly agree within their reported uncertainties. The common proper motion pair \#\,1145 is not shown in Fig.\,\ref{fig:chemistry} because its [Fe/H]~$ = -3.45$\,dex is beyond the plot ranges.

\begin{figure*}
    \centering
    \includegraphics[width=0.9\linewidth]{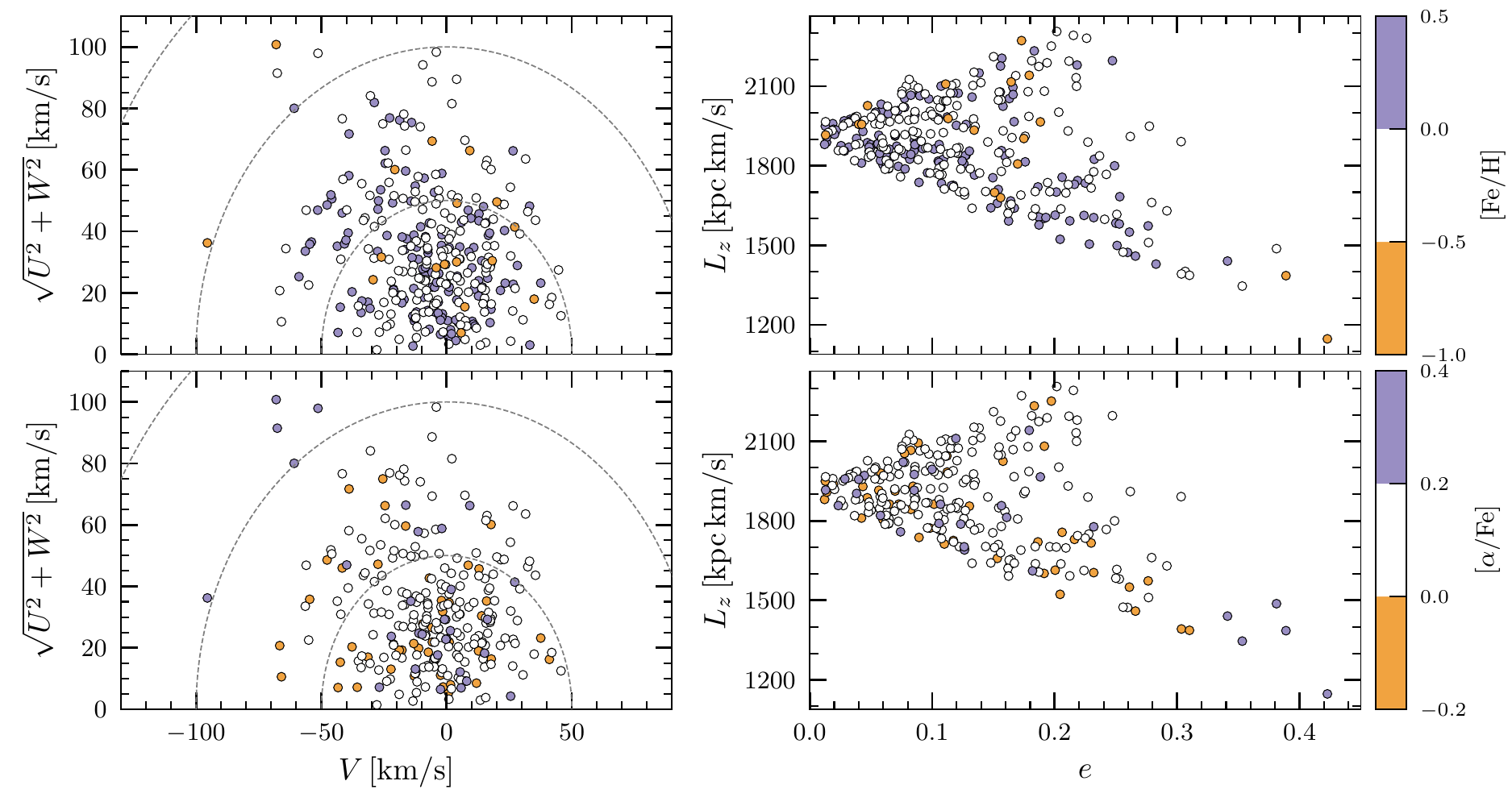}
    \caption{Toomre diagram (left panels) and azimuthal action versus eccentricity (right panels) of the common proper motion pairs that are color-coded according to the measured [Fe/H] and [$\alpha$/Fe] of the nondegenerate stars in the top and bottom panels, respectively. The adopted color scheme is shown on the right.}
    \label{fig:kinematics_feh-alpha}
\end{figure*}
\begin{figure*}
    \centering
    \includegraphics[width=0.9\linewidth]{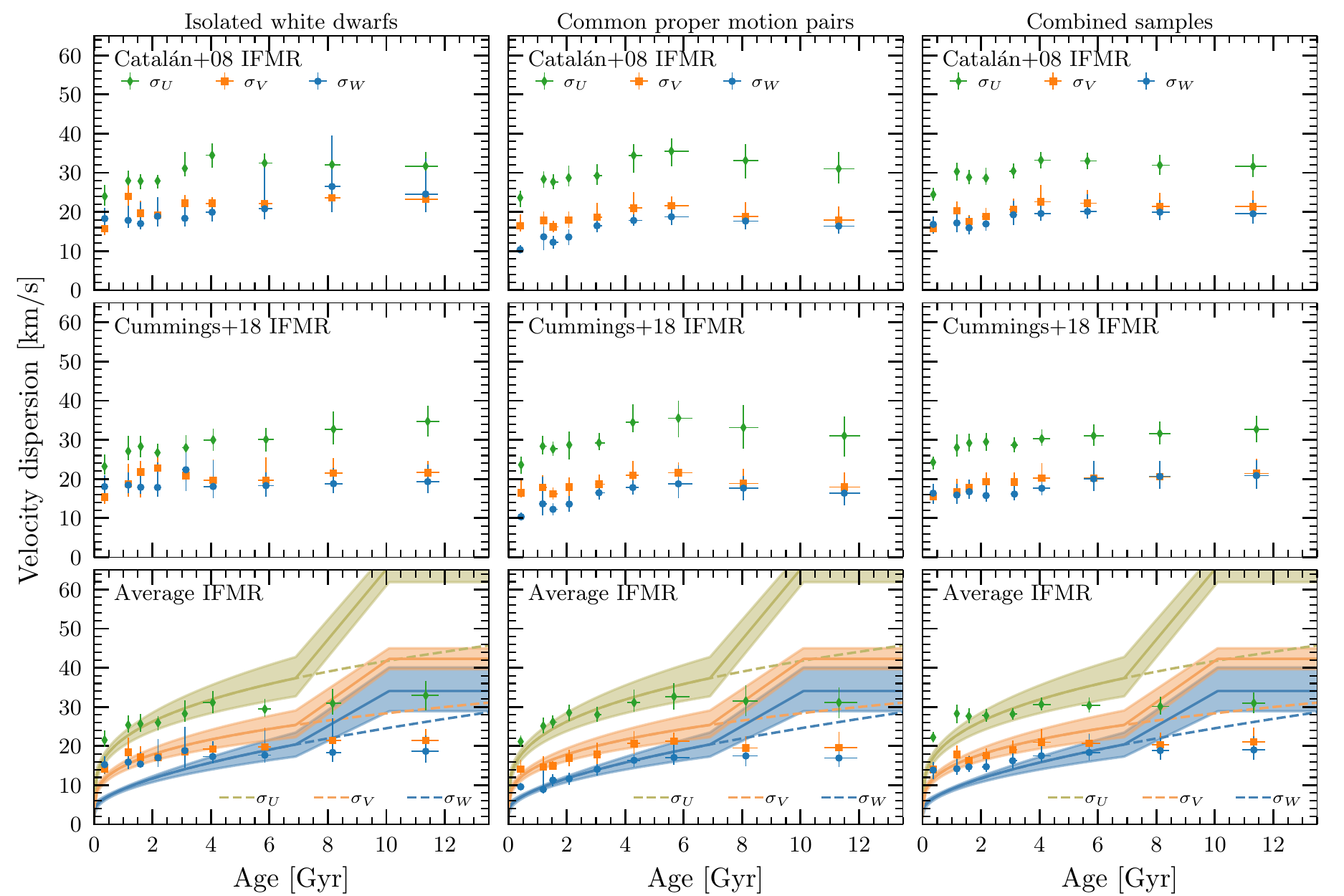}
    \caption{Age -- velocity dispersion relations for the isolated white dwarfs (left panels), the white dwarfs in common proper motion pairs (middle panels), and the combined samples (right panels). Each row shows the results obtained by considering two white dwarf progenitor age estimates, which are based on the progenitor masses that we derived via the \citet{catalan2008b} and \citet{cummings2018} IFMRs, or the average of the two. The three velocity components use different symbols and colors, as is shown in the legend of the top panels. The shaded regions in the bottom panels represent the 16--84\% ranges of the \citet{cheng2019} formulation for the age -- velocity dispersion of white dwarfs in the Galactic thin and thick disk. The dashed curves represent the extrapolated thin-disk contribution. }
    \label{fig:velocity-disp}
\end{figure*}

The spatial distribution of these stars is shown in Fig.\, \ref{fig:xyz_feh-alpha}, in which we have color-coded them according to the measured [Fe/H] and [$\alpha$/Fe] of the nondegenerate stars. The spatial distribution of this sub-sample and the median distance from the Sun are roughly the same as those of the entire sample of common proper motion pairs. We note a spatial bias towards $Z < 0$ and $\sqrt{X^2+Y^2}< R_0$ for stars of [$\alpha$/Fe]~$>0.2$\,dex that is due to the RAVE sample, which was observed from the Australian Astronomical Observatory in the Southern Hemisphere. The overall lack of correlation between metallicity and coordinates that is observed here is also seen in much larger samples of stars and is linked to the absence of a well defined age-metallicity relation \citep{casagrande20111,bergmann2014}, which also characterizes the binary sample analyzed by \citet{rebassa-mansergas2021}. Both aspects are related to the radial mixing and churning that take place in galaxies, where stars of different composition travel far away from their birth sites \citep{schonrich2009, minchev2018,hayden2018}.

As for the full sample, we also show the Toomre diagram and the azimuthal action versus eccentricity plots in Fig.\,\ref{fig:kinematics_feh-alpha}. Even here we do not see any straightforward age versus kinematics relation, but we note two interesting  aspects. First, the majority of the systems have $e<0.3$, as is typical for a sample dominated by low-$\alpha$ abundances ([$\alpha$/Fe]\,$<0.2$\,dex). Second, there are five $\alpha$-enriched systems that have $e > 0.3$, which could qualify them as members of the thick disk. Four of these systems (2632,
2653, 2655, and 3089) have $-0.7 < \mathrm{[Fe/H]} < -0.3$, while 2873 has [Fe/H]~$= 0.18$\,dex. The Galactic orbits of these five star reach $R_{\rm peri}=$~3.5--4.5\,kpc and have apocenters that do not go beyond $10$\,kpc. The very metal poor star in our sample, number 1145, has an unusual thin disk kinematics with a basically circular Galactic orbit.

\subsection{Age -- velocity dispersion relations}
We investigated the dynamic properties of the studied sample further by constructing a set of age -- velocity dispersion relations that are shown in Fig.\,\ref{fig:velocity-disp}. Each panel separately shows the age -- velocity dispersion for the isolated white dwarfs, the common proper motion pairs, and the combined sample. The age -- velocity dispersion relations are considered separately for the \citet{catalan2008b} and \citet{cummings2018} IFMRs and for the average of the two. To compute the data points shown in the figure, we only considered systems that have $e<0.27$, which we adopted as cut off for the thin disk. We removed dynamically classified thick-disk and halo members because their age uncertainties overlap with that of the thin-disk candidates, and they may therefore affect our results. We also excluded isolated white dwarfs from the SPY sample whose photometric $T_{\rm eff}$ differs by more than 30\,\% from the spectroscopic measurement. The age -- velocity dispersion relations were computed by sampling the white dwarf total ages ($ = \tau_{\rm cool}+\tau_{\rm prog}$) with a Monte Carlo method, accounting for the measured uncertainties. For the systems of unknown spectral type, we took a 2/8 ratio of non-DA/DA atmospheres into account by randomly drawing ages from the two estimates given in Tables\,\ref{tab:parameters} and \ref{tab:parameters_db}. When the total ages were constrained by lower limits in these tables, we only included the fraction of the derived distributions that are below 13.5\,Gyr, which we assumed to be the total age of the Milky Way. We computed the velocity dispersion for each randomly drawn sample by dividing the stars in nine logarithmically spaced age bins (0, 1, 1.4, 1.9, 2.6, 3.7, 5.1, 7, 9.7, 13.5\,Gyr). We computed the median and the 16--84\,\% uncertainties in each age bin in this way after initially removing the 2-$\sigma$ outliers and requiring at least 35 stars per age bin. We included an age bin older than 10\,Gyr for computational purposes and because the age distribution of our thin disk candidates extends beyond the typical estimate for the Galactic disk. 
\begin{figure}
    \centering
    \includegraphics[width=0.9\linewidth]{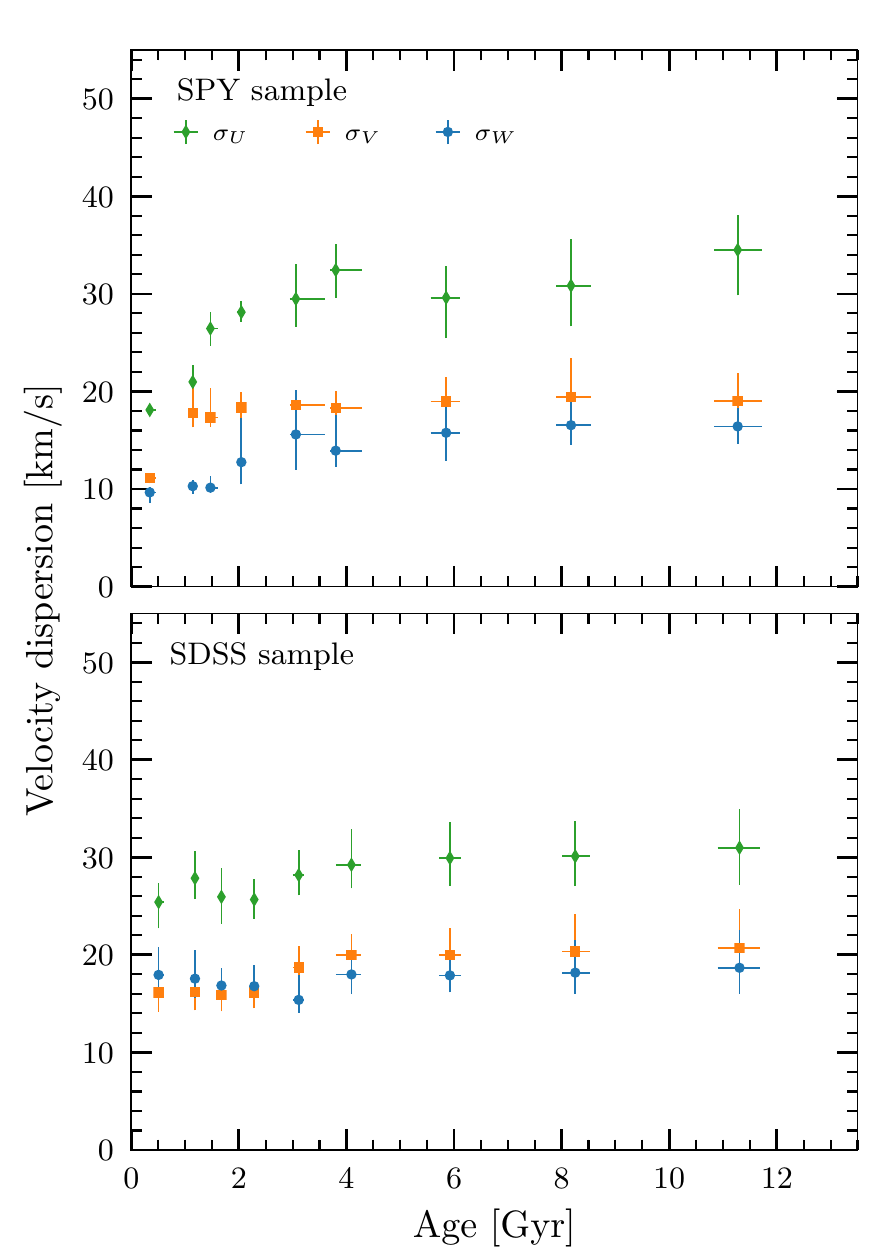}
    \caption{Age -- velocity dispersion relations for the isolated white dwarf sub-samples. }
    \label{fig:velocity-disp_wds}
\end{figure}
\begin{figure}
    \centering
    \includegraphics[width=0.9\linewidth]{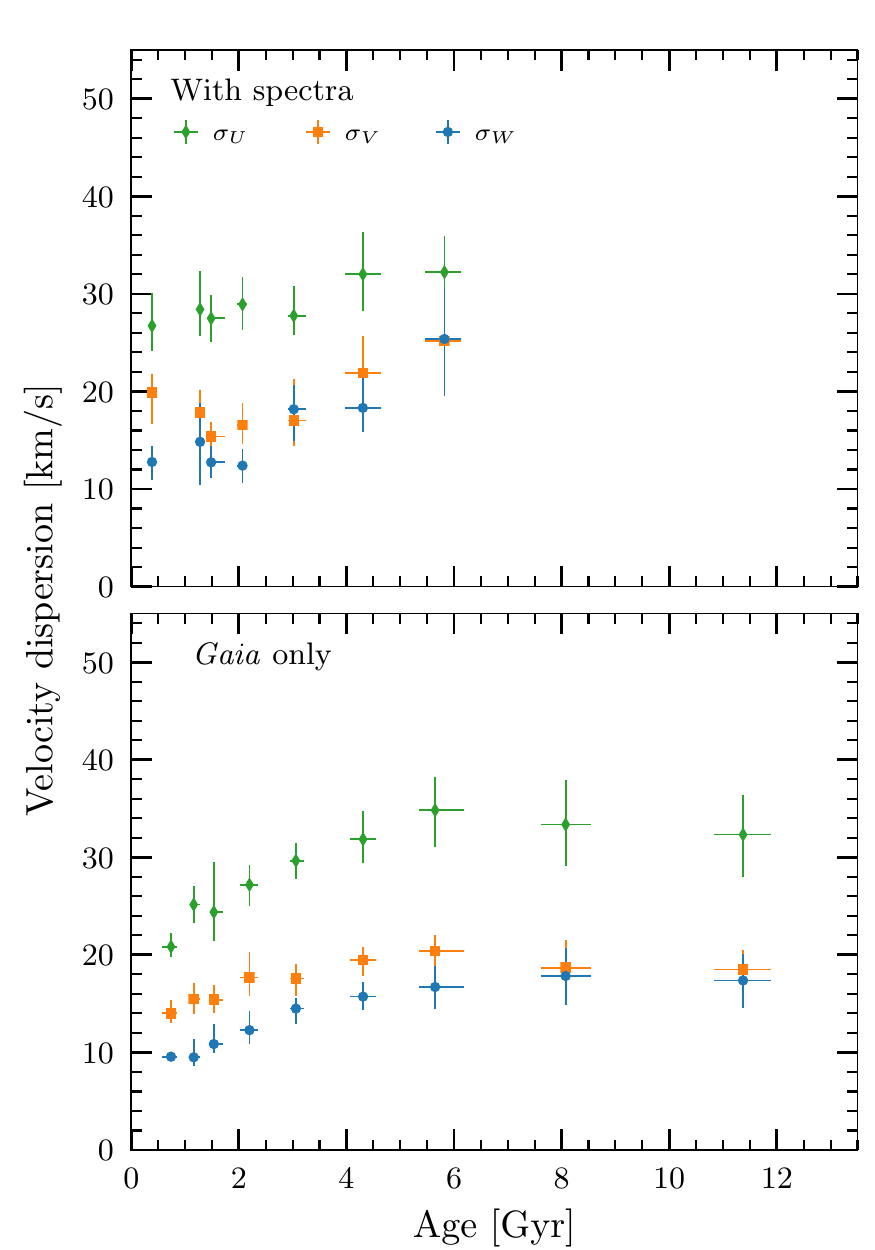}
    \caption{As Fig.\,\ref{fig:velocity-disp_wds}, but showing the age -- velocity dispersion relations for the white dwarfs in common proper motion pairs that have either observations from spectroscopic surveys or {\em Gaia}-DR2 radial velocities.}
    \label{fig:velocity-disp_cpmp}
\end{figure}

Inspecting Fig.\,\ref{fig:velocity-disp} from left to right, we note a general increase in velocity dispersion up to 6\,Gyr and a saturation that coincides with relatively flatter  velocity dispersion at larger ages. The age -- velocity dispersion relation of isolated white dwarfs that uses the \citet{catalan2008b} IFMR shows a slight decrease in the last bins, in contrast to that using the \citet{cummings2018} IFMR. This difference is likely caused by the larger number of white dwarfs of lower progenitor mass that are obtained with the latter IFMR, which in turn have longer total ages. The observed trends are typically more evident for the $\sigma_U$ component, while the $\sigma_V$ and $\sigma_W$ components are noisier, especially in the first few age bins of the isolated white dwarf sample. The three bottom panels of Fig.\,\ref{fig:velocity-disp} show a comparison of our averaged results with the parameterization obtained by \citet{cheng2019}, who modeled thin- and thick-disk contributions by analyzing a {\em Gaia}-selected white dwarf sample within 250\,pc and using FGK-type stars as reference. Our results are obtained averaging the $\tau_{\rm prog}$, which were inferred from the progenitor masses obtained with the two considered IFMRs. The $\sigma_U$ component of the age -- velocity dispersion relation for the isolated white dwarfs follows the \citet{cheng2019} formulation up to 4\,Gyr, but it continues on a relatively straight line at older ages. The $\sigma_V$ and $\sigma_W$ components, instead, are more or less flat throughout the entire age range. The best agreement is obtained for the age -- velocity dispersion relation of the common proper motion pairs, which closely follows the \citet{cheng2019} formulation up to 6-7\,Gyr and reaches a flat asymptote for older ages. The asymptotic values are below the extrapolated curves for the thin disk contribution to the age -- velocity dispersion relation of \citet{cheng2019}, where these authors impose a dominant contribution from the thick disk for ages $>7$\,Gyr, while we excluded the kinematically selected thick-disk and halo members. The age -- velocity dispersion of the two combined samples is more strongly dominated by the isolated white dwarf sample, and it therefore compares less well with the \citet{cheng2019} formulation, especially in the $\sigma_W$ component. The three average relations are tabulated in Table\,\ref{tab:avd_uber_relation}.
\begin{figure*}
    \centering
    \includegraphics[width=0.9\linewidth]{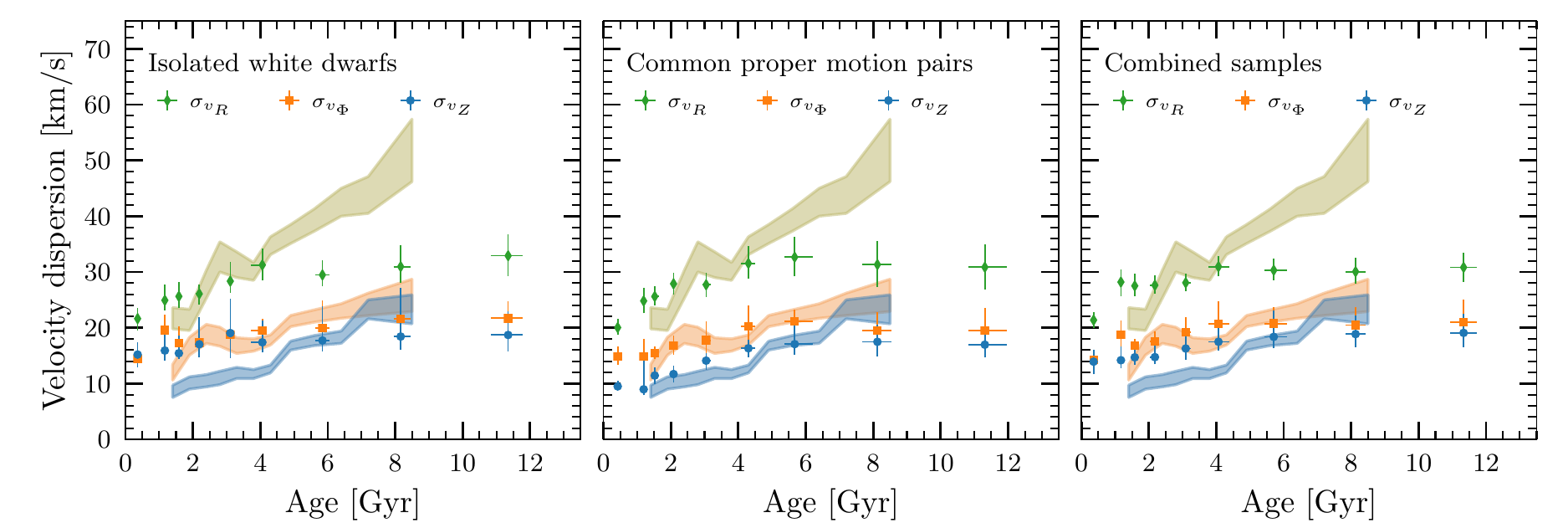}
    \caption{Age -- velocity dispersion relations in Galactic cylindrical coordinates. Our results (data points with error bars) are compared to the results of \citet{yu2018} for LAMOST FGK-type stars with [Fe/H]~$-0.2$\,dex. The colored bands represent their 1-$\sigma$ error ranges.}
    \label{fig:velocity-disp_cyl}
\end{figure*}
\begin{table*}
    \centering
    \caption{Average age -- velocity dispersion relations that are shown in the bottom panels of Fig.\,\ref{fig:velocity-disp}.}
    \scriptsize
    \begin{tabular}{@{}cccccccccccc@{}}
        \toprule
        \toprule
    Age bin\,[Gyr]  & $\sigma_{U}$\,[km\,s$^{-1}$] & $\sigma_{V}$\,[km\,s$^{-1}$] & $\sigma_{W}$\,[km\,s$^{-1}$] & 
    Age bin\,[Gyr]  & $\sigma_{U}$\,[km\,s$^{-1}$] & $\sigma_{V}$\,[km\,s$^{-1}$] & $\sigma_{W}$\,[km\,s$^{-1}$] &    
    Age bin\,[Gyr]  & $\sigma_{U}$\,[km\,s$^{-1}$] & $\sigma_{V}$\,[km\,s$^{-1}$] & $\sigma_{W}$\,[km\,s$^{-1}$] \\    
\midrule    
\multicolumn{4}{c}{Isolated white dwarfs} & \multicolumn{4}{c}{Common proper motion pairs} & \multicolumn{4}{c}{Full sample}\\
\midrule 
$ 0.36^{+0.05}_{-0.04} $ & $21.53^{+2.41}_{-1.69} $ &$ 14.19^{+0.83}_{-1.37} $ & $15.25^{+2.10}_{-2.31} $ &$ 0.42^{+0.04}_{-0.04} $ & $21.15^{+1.52}_{-1.40} $ &$ 14.07^{+2.44}_{-1.40} $ & $9.59^{+0.97}_{-0.68} $ &$ 0.36^{+0.08}_{-0.04} $ & $22.21^{+1.34}_{-1.20} $ &$ 14.03^{+0.99}_{-1.05} $ & $13.87^{+2.12}_{-2.29} $ \\
$ 1.17^{+0.03}_{-0.03} $ & $25.37^{+2.48}_{-1.95} $ &$ 18.39^{+3.77}_{-3.77} $ & $15.90^{+3.09}_{-1.82} $ &$ 1.19^{+0.09}_{-0.07} $ & $25.13^{+2.11}_{-1.88} $ &$ 14.69^{+2.68}_{-1.48} $ & $8.92^{+5.05}_{-1.04} $ &$ 1.17^{+0.07}_{-0.03} $ & $28.29^{+2.24}_{-2.39} $ &$ 17.84^{+2.28}_{-2.41} $ & $14.19^{+2.43}_{-1.49} $ \\
$ 1.59^{+0.05}_{-0.04} $ & $25.69^{+2.46}_{-2.08} $ &$ 17.09^{+2.78}_{-1.75} $ & $15.40^{+3.09}_{-1.00} $ &$ 1.52^{+0.13}_{-0.07} $ & $26.12^{+1.73}_{-1.60} $ &$ 14.92^{+1.21}_{-0.96} $ & $11.30^{+1.51}_{-1.61} $ &$ 1.58^{+0.04}_{-0.11} $ & $27.63^{+2.00}_{-1.72} $ &$ 16.27^{+1.38}_{-1.09} $ & $14.69^{+2.65}_{-1.32} $ \\
$ 2.19^{+0.03}_{-0.12} $ & $25.97^{+1.64}_{-2.01} $ &$ 17.13^{+1.61}_{-1.29} $ & $17.01^{+4.84}_{-2.37} $ &$ 2.08^{+0.10}_{-0.09} $ & $28.52^{+2.01}_{-2.07} $ &$ 16.94^{+2.03}_{-2.04} $ & $11.64^{+1.53}_{-1.54} $ &$ 2.18^{+0.03}_{-0.14} $ & $27.78^{+1.77}_{-1.51} $ &$ 17.55^{+1.81}_{-1.59} $ & $14.72^{+2.32}_{-1.29} $ \\
$ 3.11^{+0.11}_{-0.09} $ & $28.28^{+3.47}_{-2.12} $ &$ 18.51^{+1.93}_{-1.54} $ & $18.84^{+6.09}_{-4.34} $ &$ 3.04^{+0.15}_{-0.10} $ & $28.02^{+2.02}_{-1.74} $ &$ 17.88^{+2.99}_{-2.89} $ & $14.07^{+1.35}_{-1.67} $ &$ 3.08^{+0.15}_{-0.09} $ & $28.17^{+1.57}_{-1.43} $ &$ 19.06^{+2.41}_{-2.03} $ & $16.25^{+3.63}_{-1.97} $ \\
$ 4.06^{+0.08}_{-0.33} $ & $31.12^{+2.93}_{-2.77} $ &$ 19.26^{+2.05}_{-1.57} $ & $17.33^{+3.96}_{-1.81} $ &$ 4.30^{+0.20}_{-0.23} $ & $31.13^{+3.24}_{-2.28} $ &$ 20.65^{+3.18}_{-2.68} $ & $16.36^{+1.49}_{-1.66} $ &$ 4.06^{+0.32}_{-0.29} $ & $30.65^{+1.85}_{-1.58} $ &$ 20.96^{+3.34}_{-2.45} $ & $17.52^{+2.49}_{-1.57} $ \\
$ 5.84^{+0.20}_{-0.23} $ & $29.43^{+2.59}_{-2.01} $ &$ 19.73^{+4.93}_{-1.74} $ & $17.67^{+6.37}_{-1.98} $ &$ 5.67^{+0.52}_{-0.30} $ & $32.63^{+3.54}_{-3.24} $ &$ 21.23^{+2.15}_{-2.31} $ & $17.09^{+3.55}_{-1.91} $ &$ 5.70^{+0.38}_{-0.26} $ & $30.36^{+2.02}_{-1.71} $ &$ 20.67^{+2.64}_{-1.87} $ & $18.37^{+4.65}_{-2.06} $ \\
$ 8.16^{+0.29}_{-0.22} $ & $30.90^{+3.74}_{-2.91} $ &$ 21.43^{+4.36}_{-2.93} $ & $18.37^{+8.73}_{-2.37} $ &$ 8.13^{+0.41}_{-0.46} $ & $31.50^{+4.04}_{-3.81} $ &$ 19.51^{+3.09}_{-2.80} $ & $17.48^{+2.96}_{-2.56} $ &$ 8.13^{+0.29}_{-0.28} $ & $30.13^{+2.40}_{-2.09} $ &$ 20.37^{+2.99}_{-2.22} $ & $18.84^{+3.37}_{-2.31} $ \\
$ 11.35^{+0.43}_{-0.50} $ & $32.93^{+3.69}_{-3.72} $ &$ 21.48^{+2.97}_{-2.12} $ & $18.66^{+4.14}_{-2.91} $ &$ 11.31^{+0.63}_{-0.49} $ & $31.13^{+3.86}_{-4.02} $ &$ 19.60^{+3.98}_{-3.09} $ & $16.94^{+2.53}_{-2.34} $ &$ 11.34^{+0.39}_{-0.40} $ & $30.96^{+2.66}_{-2.64} $ &$ 21.02^{+3.80}_{-2.52} $ & $19.00^{+3.45}_{-2.52} $ \\
\bottomrule
    \end{tabular}
    \label{tab:avd_uber_relation}
\end{table*}

In Fig.\,\ref{fig:velocity-disp_wds} and \ref{fig:velocity-disp_cpmp}, we break down the contribution of each sub-sample to the age -- velocity dispersion relations. Although the SPY sample contains fewer objects than the SDSS sample (top and bottom panels of Fig.\ref{fig:velocity-disp_wds}), it clearly shows a rise of the  velocity dispersion with increasing age and a saturation that begins at $\sim$4\,Gyr. The SDSS sample, as noted by \citet{anguiano2017}, has a much flatter age -- velocity dispersion relation at almost all ages, which the authors interpreted as being caused by an unidentified source of dynamical heating in addition to that caused by the secular evolution of the Galactic disk. The still possibly inaccurate radial velocity measurements of the SDSS sample, despite our calibration obtained in Sect.~\ref{sec:sample_wd}, may be the cause of the disagreement between the SPY and SDSS samples. However, we note that the SPY sample was cleaned of objects that had discrepant $T_{\rm eff}$ measurements in between the spectroscopic and photometric analyses, while the SDSS sample included all stars initially selected by us in Sect.~\ref{sec:sample_wd}. Nevertheless, we also note that the maximum  velocity dispersion we have measured for both samples agree with their averages, which were previously determined for the thin-disk stars in the SPY sample \citep{pauli2006}. The common proper motion pairs with spectra comprise just 314 stars, which limits our ability to determine a meaningful age -- velocity dispersion relation beyond 6\,Gyr (top panel of Fig.\,\ref{fig:velocity-disp_cpmp}). This result is quite noisy, especially in the earliest age bins. The age -- velocity dispersion relation of the common proper motion pairs that only have {\em Gaia} DR2 radial velocities features a smoother increase. Furthermore, we note that the three components clearly reach an asymptotic behavior in the range of 15--35\,km\,s$^{-1}$ at $\sim$6\,Gyr, which is expected to be caused by dynamical heating of stars as they interact within the Galactic disk potential \citep[cf.][and references therein]{seabroke2007}. The values we measured for the age -- velocity dispersion relation appear to be smaller than those obtained by previous work on wide binaries \citep{wegner1981, silvestri2001}, which considered much smaller samples without separating stars in age bins or eccentricity ranges, however. They therefore likely included at least some thick-disk members.

We also computed the age -- velocity dispersion relation in Galactic cylindrical coordinates, which we compare in  Fig.\,\ref{fig:velocity-disp_cyl} to the age -- velocity dispersion relation of LAMOST FGK-type stars with [Fe/H]~$>-0.2$  \citep{yu2018}. The studied samples follow the general increasing trend observed for the LAMOST stars up to 4--6\,Gyr, where our age -- velocity dispersion relations reach saturation. On the other hand, the LAMOST selection, which favors $\alpha$-depleted stars that are more likely thin-disk members, shows a $\sigma_{U}$ component that rises up to 50\,km\,s$^{-1}$.

\begin{table}
    \centering
    \caption{Velocity dispersion as function of [Fe/H] and [$\alpha$/Fe].}
    \begin{tabular}{@{}cccc@{}}
    \toprule
    \toprule
$[{\rm Fe/H}]$ & $\sigma_{U}$\,[km\,s$^{-1}$] & $\sigma_{V}$\,[km\,s$^{-1}$] & $\sigma_{W}$\,[km\,s$^{-1}$]  \\
    \midrule
$ (-1.00,  -0.50) $ & $ 42.73_{-2.12}^{+2.06} $ & $26.24_{-5.74}^{+3.50} $ & $20.37_{-0.51}^{+1.42} $\\
$ (-0.50,  0.00) $ & $ 31.55_{-1.15}^{+0.73} $ & $20.52_{-1.16}^{+0.65} $ & $19.88_{-3.54}^{+2.14} $\\
$ (0.00, 0.60) $ & $ 28.29_{-1.35}^{+1.50} $ & $22.27_{-1.31}^{+1.08} $ & $18.40_{-0.77}^{+1.45} $\\
   
    \toprule
$[{\rm \alpha/Fe}]$ & $\sigma_{U}$\,[km\,s$^{-1}$] & $\sigma_{V}$\,[km\,s$^{-1}$] & $\sigma_{W}$\,[km\,s$^{-1}$]  \\
    \midrule
$ (-0.20,  0.00) $ & $ 31.59_{-2.69}^{+2.75} $ & $24.73_{-1.67}^{+1.93} $ & $15.22_{-1.21}^{+1.67} $\\
$ (0.00,  0.20) $ & $ 30.82_{-0.91}^{+1.66} $ & $19.79_{-1.23}^{+2.29} $ & $18.28_{-3.75}^{+4.12} $\\
$ (0.20,  0.60) $ & $ 30.83_{-2.56}^{+2.50} $ & $19.63_{-2.45}^{+2.78} $ & $18.14_{-2.98}^{+3.17} $\\

\bottomrule
            \end{tabular}
    \label{tab:avd_relation_metals}
\end{table}
To conclude this section, we measured the velocity dispersion of the common proper motion pairs by binning their [Fe/H] and [$\alpha$/Fe]. The results of this exercise are given in Table\,\ref{tab:avd_relation_metals}. Here, even taking into account the small size of our sample, we note that the radial component, $\sigma_{U}$, has the largest dispersion in the case of the most metal-poor systems and is anticorrelated with increasing metallicity, while the other two components have the same  velocity dispersion within the errors.  The observed trend for $\sigma_U$ may be related to the radial mixing occurring in the Milky Way, as discussed in the previous sections. The three  velocity dispersion components do not show a significant trend with [$\alpha$/Fe]. The average dispersions of each velocity component are in agreement with those from much larger samples of single stars \citep[cf.][]{hayden2020}.

\section{Summary and outlook}
We have analyzed, for the first time, the 3D kinematics of 3133 isolated white dwarfs with available radial velocity measurements and white dwarfs in common proper motion pairs that contain one nondegenerate companion with radial velocity measurements. The studied stars have highly reliable {\em Gaia} EDR3 data, and in about 30\% of the common proper motion pairs, measured metal abundances for the nondegenerate stars. We uniformly estimated the white dwarf physical parameters (mass and cooling age) by interpolation with state of the art evolutionary sequences in the {\em Gaia} HR diagram. Using a set of initial-to-final-mass relations and main-sequence evolutionary tracks, we estimated the white dwarf progenitor masses and total ages. This approach has enabled us to identify kinematic members of the Galactic thin and thick disk, and of the halo. The latter group contains seven isolated white dwarfs. Another previously classified halo member is confirmed to belong to the thin disk, thanks to the accurate proper motions of {\em Gaia} EDR3. We have measured the age -- velocity dispersion relation for the studied sample, breaking it down by sub-samples, in agreement with previous results that used different methods. Moreover, our age -- velocity dispersion relations show both signs of dynamical heating and saturation beginning at 4--6\,Gyr. Taking advantage of the measured abundances for the nondegenerate companions in common proper motion pairs, we confirmed the presence of overlapping populations in the solar neighborhood, for instance, due to radial mixing, and we measured an anticorrelation between [Fe/H] and the  velocity dispersion along the radial direction in the Galactic reference frame. The studied sample can be further exploited to improve classification schemes of white dwarfs belonging to the thin and thick disk and to the halo, and also to study their membership to local streams and moving groups \citep[][]{torres2019b}.

In the coming years, the legacy of the existing large spectroscopic surveys such as RAVE, LAMOST, APOGEE, and GALAH, will be taken upon by much larger projects as WEAVE \citep{weave}, 4MOST \citep{4most}, and SDSS-V \citep{sdssV}, that will observe at the same time all {\em Gaia} white dwarfs and millions of stars for chemical tagging of the Galactic populations. Hence, a more coherent picture of the global chemodynamics of the Milky Way and its connection with the solar neighborhood and the white dwarf population will come within reach.

\begin{acknowledgements}

RR has received funding from the postdoctoral fellowship programme Beatriu de Pin\'os, funded by the Secretary of Universities and Research (Government of Catalonia) and by the Horizon 2020 programme of research and innovation of the European Union under the Maria Sk\l{}odowska-Curie grant agreement No 801370. ARM acknowledges additional support from Grant RYC-2016-20254 funded by MCIN/AEI/10.13039/501100011033 and by ESF Investing in your future. ST acknowledges support from the MINECO under the AYA2017-86274-P grant, and the AGAUR grant SGR-661/2017. JM acknowledges  support from  the Accordo Attuativo  ASI-INAF n. 2018.22.HH.O, Partecipazione alla fase B1 della missione Ariel. MEC acknowledges NASA grants 80NSSC17K0008 and 80NSSC20K0193. PET and TC have received funding from the European Research Council under the European Union’s Horizon 2020 research and innovation programme n. 677706 (WD3D). UH acknowledges funding by the Deutsche Forschungsgemeinschaft (DFG) through grants IR190/1-1, HE1356/70-1, and HE1356/71-1. TC was supported by a Leverhulme Trust Grant (ID RPG-2020-366). JJR is thankful for support from NSFC (grant No. 11903048, 11833006).

Project supported by a 2019 Leonardo Grant for Researchers and Cultural Creators, BBVA Foundation. The Foundation accepts no responsibility for the opinions, statements and contents included in the project and/or the results thereof, which are entirely the responsibility of the authors. 

This work has made use of data from the European
Space Agency (ESA) mission {\em Gaia} (\url{https://www.cosmos.esa.int/gaia}), processed by the Gaia Data Processing and Analysis Consortium (DPAC, \url{https://www.cosmos.esa.int/web/gaia/dpac/consortium}). Funding for the DPAC has been provided by national institutions, in particular the institutions participating in the Gaia Multilateral Agreement. This research has made use of the SIMBAD database \citep{wenger2000}, operated at CDS, Strasbourg, France. This research has made use of the VizieR catalogue access tool, CDS, Strasbourg, France. The original description of the VizieR service was published in \citet{ochsenbein20000}. The TAP service used in this paper was constructed as part of the activities of the German Astrophysical Virtual Observatory. 

All figures of this manuscript have been made with the {\sc python}'s package \verb|matplotlib| \citep{matplotlib}.

 \end{acknowledgements}

% WARNING
%-------------------------------------------------------------------
% Please note that we have included the references to the file aa.dem in
% order to compile it, but we ask you to:
%
% - use BibTeX with the regular commands:
%   \bibliographystyle{aa} % style aa.bst
%   \bibliography{Yourfile} % your references Yourfile.bib
%
% - join the .bib files when you upload your source files
%-------------------------------------------------------------------

\bibliographystyle{aa}
\bibliography{kinematics2021}

\end{document}